\documentclass[12pt]{JHEP3}
\usepackage{epsfig}
\usepackage{amsmath}

\makeatletter
\@addtoreset{equation}{section}

\newcommand{\p}{\partial}
\newcommand{\nn}{\nonumber}
\newcommand{\rmv}[1]{}
\newcommand{\identity}{{\rlap{1} \hskip 1.6pt \hbox{1}}}

\title{Strings between branes} 

\author{Koji Hashimoto \\ 
Institute of Physics, University of Tokyo, Komaba
Tokyo 153-8902, Japan\\
E-mail :  \email{koji {\rm at} hep1.c.u-tokyo.ac.jp}}

\author{Washington Taylor \\ 
Center for Theoretical Physics, MIT
Bldg.\ 6-308, Cambridge, MA 02139, U.S.A\\
E-mail : \email{wati {\rm at} mit.edu}
}

\abstract{
D-brane configurations containing fundamental
strings are constructed as classical solutions of Yang-Mills theory.
The fundamental strings in these systems stretch between D-branes.  In
the case of D1-branes, this construction gives smooth (classical)
resolutions of string junctions and string networks.  Using a
non-abelian Yang-Mills analysis of the string current, the string
charge density is computed and is shown to have support in the region
between the D-brane world-volumes.  The $\mbox{'t Hooft}$-Polyakov
monopole is analyzed using similar methods, and is shown to contain
D-strings whose flux has support off the D-brane world-volume defined by
the Higgs scalar field, when this field is interpreted in terms of a
transverse dimension.  The constructions presented here are used to
give a qualitative picture of tachyon condensation in the Yang-Mills
limit, where fundamental strings and lower-dimensional D-branes arise
in a volume of space-time where brane-antibrane annihilation has
occurred.
}

\preprint{
{\normalsize\tt hep-th/0307297}\\
{\normalsize MIT-CTP-3392}\\
{\normalsize UT-Komaba/03-12}
}


\begin{document}

%%%%%%%%%%%%%%%%%%%%%%%%%%%%%%%%%%%%%%%%%%%%%%%%%%%%%%%%%%%%%%%%%%%%
%%%%%%%%%%%%%%%%%%%%%%%%%%%%%%%%%%%%%%%%%%%%%%%%%%%%%%%%%%%%%%%%%%%%
%%%%%%%%%%%%%%%%%%%%%%%%%%%%%%%%%%%%%%%%%%%%%%%%%%%%%%%%%%%%%%%%%%%%

\section{Introduction}
\label{sec1}

While the abelian world-volume theory on a single D$p$-brane has a
simple geometric interpretation in terms of a brane moving in
space-time under a Born-Infeld action \cite{Leigh,Polchinski-TASI}, the
nonabelian world-volume theory on a system of several D$p$-branes has
a much richer structure.  The off-diagonal strings on a system of $N$
D$p$-branes give rise to an enhanced U($N$) symmetry when the branes
are coincident \cite{Witten}, and can give rise to interesting
phenomena such as higher-dimensional D$(p + 2n)$-branes when the
branes are separated \cite{Taylor-Trieste}, such as by the application
of an external field \cite{Myers}.  While the full nonabelian
generalization of Born-Infeld theory describing the dynamics of the
massless fields on $N$ D$p$-branes is not fully understood, in the
low-energy limit this theory reduces to U($N$) maximally
supersymmetric $(p + 1)$-dimensional Yang-Mills theory.  Many
interesting features of the nonabelian brane system are captured, at
least qualitatively, by the low-energy Yang-Mills theory on the
D$p$-branes.

In this paper we consider a novel construction in U($N$) Yang-Mills
theory.  We find a family of classical solutions of the theory which
represent BPS configurations of multiple D$p$-branes with strings
stretched between them.  For D1-branes, in particular, this
construction gives a representation in classical Yang-Mills theory of
string junctions and networks \cite{das,ggt,Sen-network}.  We analyze
the spatial distribution of the string current in these configurations
using the approach to nonabelian current analysis developed in
\cite{WT-Mark,Myers,Okawa-Ooguri}.  We find that the string current
flows through the region between the D-branes, so that the Yang-Mills
theory is actually capable of describing physics in regions of
space-time away from the world-volume of the branes.  This feature
differentiates the construction presented here from related
configurations in the abelian theory such as the BIon
\cite{CM,hlw,Gibbons}, where electric flux corresponding to string
charge lives on the D-brane world-volume, which may itself contain
spikes or other complex geometry associated with string/brane
intersection configurations.

As another example of the methods developed in this paper, we
reconsider the 't Hooft-Polyakov monopole.  As was discussed in
\cite{Aki}, when the Higgs scalar in the Prasad-Sommerfield SU(2)
monopole solution \cite{Prasad-Sommerfield} is interpreted as a
transverse scalar for a two D3-brane configuration, a natural
geometric interpretation of this monopole solution arises.  We show
that in this geometrical picture, the D-string stretching between the
two D3-branes is described by a D-string flux moving from one brane to
another.  This D-string flux leaves the world-volume of the D3-branes,
just as the fundamental string does in the other examples we consider.

As an application of the methods used here, we discuss the tachyon
condensation story \cite{conjectures} in Yang-Mills language.  The
basic tachyon condensation picture was related to a Yang-Mills
description of intersecting branes in \cite{Aki-WT,Morosov,HN}, where
it was shown that the tachyonic mode on the intersecting branes gives
rise to a recombination of the branes.  Putting electric flux on the
D-branes, we find a qualitative description of tachyon condensation
where intersecting branes recombine, giving a region of empty space
between the branes in which the electric flux is replaced by
fundamental strings stretching through the vacuum.  Because we are
working in the classical Yang-Mills picture, the fundamental string
charge is not quantized and the confinement of these strings is not
completely apparent; as the recombined D-branes move apart, we are left with a
distribution of stretched strings filling the vacuum.  The
related configuration in which D3-branes annihilate to give D-strings
in the vacuum manifests the brane descent relations suggested by Sen.

In Section 2 we present a new family of Yang-Mills solutions
representing D-strings with electric flux/fundamental strings
connecting them, including generalizations of the basic two D-brane
configuration to a large family of string junctions and networks.  In
Section 3 we analyze the D-string flux of the 't Hooft-Polyakov
monopole.  In Section 4 we discuss the application to tachyon
condensation.  Section 5 contains conclusions and a discussion.

%%%%%%%%%%%%%%%%%%%%%%%%%%%%%%%%%%%%%%%%%%%%%%%%%%%%%%%%%%%%%%%%%%%%
%%%%%%%%%%%%%%%%%%%%%%%%%%%%%%%%%%%%%%%%%%%%%%%%%%%%%%%%%%%%%%%%%%%%
%%%%%%%%%%%%%%%%%%%%%%%%%%%%%%%%%%%%%%%%%%%%%%%%%%%%%%%%%%%%%%%%%%%%
\section{String junctions and networks in Yang-Mills theory}
\label{sec2}

In this section we construct BPS configurations of D1-branes and
fundamental strings in Yang-Mills language.  In subsection
\ref{sec:solution} we give a simple example of a solution of the
Yang-Mills equations of motion, which describes a smooth deformation
of a pair of intersecting D1-branes with electric flux corresponding
to fundamental string charge.  In the deformed configuration the
D1-branes have reconnected and are spatially separated.  In
\ref{sec:charge} 
we analyze the spatial distribution of fundamental string charge, and
show that it contains a part which flows through the region of space
between the D1-branes.  Subsection \ref{sec:junction} relates the
configuration we have constructed to a smooth family of string junctions.
Subsection \ref{sec:generalization} contains a
discussion of more general string junctions and networks which can be
described in a similar fashion using Yang-Mills theory.  

\subsection{A simple BPS solution of Yang-Mills theory}
\label{sec:solution}

We begin by describing a simple classical solution of Yang-Mills theory.
We work in (1+1)-dimensional SU(2)  Yang-Mills theory
\begin{eqnarray}
 {\cal L} & = & -T_{\rm D1} {\rm Tr}\left[
\frac14 (2\pi\alpha')^2(F_{\mu\nu})^2 +\frac12 (D_\mu Y)^2
\right]
\label{eq:l}
\end{eqnarray}
which is the low-energy effective Lagrangian describing two parallel
D1-branes.  
The indices in (\ref{eq:l}) take values $\mu,\nu=0,1$.
The field $Y$ is the scalar field describing transverse
fluctuations. Taking the gauge $A_1=0$, the 
BPS equation \cite{CM,das}
\begin{eqnarray}
 2\pi\alpha' F_{01}+ D_1 Y=0
\label{eq:bps}
\end{eqnarray}
can be solved by 
\begin{eqnarray}
 Y = 2\pi\alpha' A_0 \ .
\label{eq:ya}
\end{eqnarray}
In this case, the equations of motion reduce to the simple equation 
\begin{eqnarray}
(\partial_1)^2 Y=0 \ ,
\label{eq:y-equation}
\end{eqnarray}
if we assume that $Y$ has no time dependence.

\begin{figure}[tp]
\begin{center}
\begin{minipage}{10cm}
\begin{center}
\includegraphics[width=6cm]{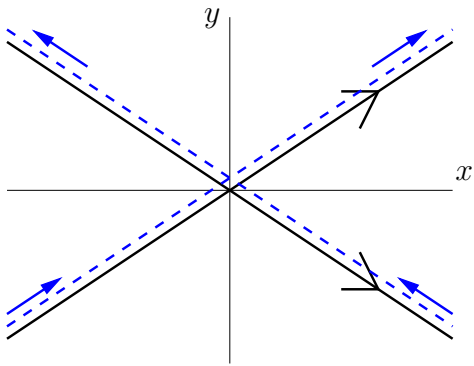}
\put(-95,130){$y$}
\put(0,70){$x$}
\caption{Intersecting $(\pm p,1)$-strings.  Arrows denote the
 orientation of the (solid) D-strings as well as the directions
 of the (dashed) fundamental string fluxes (the electric fluxes
 on the D-strings).}
\label{juncorigfig}
\end{center}
\end{minipage}
\end{center}
\end{figure}

Let us consider the classical solution to
(\ref{eq:y-equation}) given by
\begin{eqnarray}
 Y = \left(
\begin{array}{cc}
px & a \\ a & -px
\end{array}
\right)
\label{sol}
\end{eqnarray}
where $a$ and $p$ are positive constant parameters, and 
$x \equiv x_1$.
Since the field $Y$ measures the transverse displacement of the
D-strings,  when $a = 0$
we can interpret the geometry of this configuration as a pair of
intersecting D-strings with slopes
\begin{eqnarray}
\pm p = \pm \tan (\theta/2)
\end{eqnarray}
where $\theta$ ($0\leq \theta \leq \pi$) is the angle between the two
D-strings (See Fig.\ \ref{juncorigfig}).  Since only diagonal entries
are turned on in this case, the solution is basically given by two
independent solutions of the abelian system.  From the relation
(\ref{eq:ya}) we see that the electric flux on the two D-strings is
given by $F^{01} = \pm px/2 \pi \alpha'$.  Thus, the first part of the
solution is a $(p,1)$-string which is tilted as $Y=px$, while the
second part is a $(-p,1)$-string tilted as $Y=-px$. Since these two
satisfy the same BPS equation $2\pi\alpha'F_{01} + \partial_1 Y=0$,
the brane configuration is supersymmetric.

\begin{figure}[tp]
\begin{center}
\begin{minipage}{10cm}
\begin{center}
\includegraphics[width=9cm]{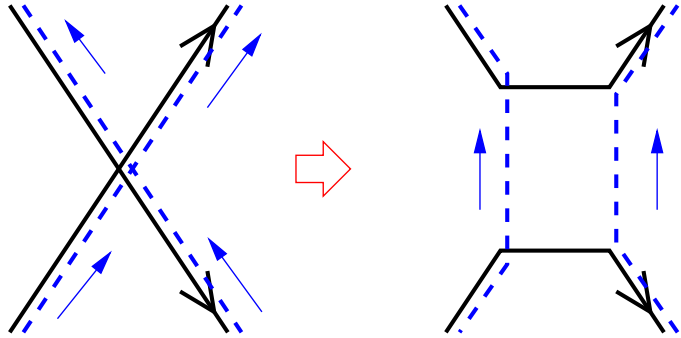}
\caption{
Generation of a box. This deformation does not affect the
 preserved supersymmetries.}
\label{boxfig}
\end{center}
\end{minipage}
\end{center}
\end{figure}

\begin{figure}[bp]
\begin{center}
\begin{minipage}{13cm}
\begin{center}
\includegraphics[width=12cm]{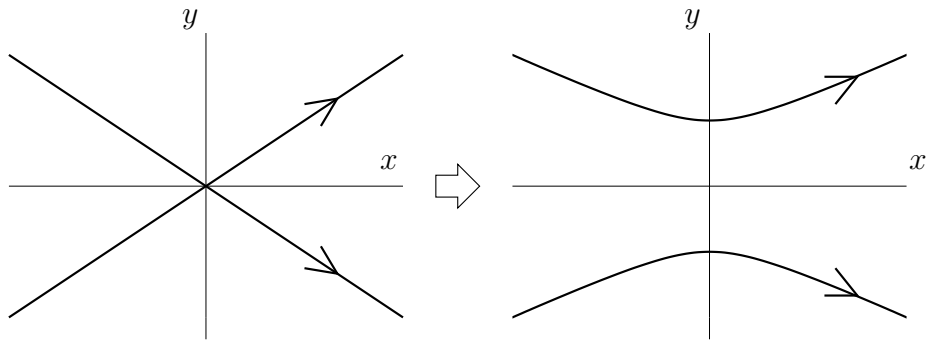}
\put(-85,120){$y$}
\put(0,65){$x$}
\put(-275,120){$y$}
\put(-200,65){$x$}
\caption{
Intersecting D-strings are recombined when the off-diagonal
  mode is turned on.}
\label{nofluxfig}
\end{center}
\end{minipage}
\end{center}
\end{figure}

It is known that this kind of supersymmetric intersection admits a
deformation which preserves supersymmetry.  Such a deformation can be
found by producing a ``box''-shaped configuration using supersymmetric
string junctions \cite{kol,AHK,Sen-network} (see Fig.\ \ref{boxfig}).
Because the parameter $a$ does not
violate the BPS condition (\ref{eq:bps}), we expect that this
parameter should provide an analogous deformation in the Yang-Mills
picture.  In fact, as shown in \cite{HN}, turning on this off-diagonal
entry in $Y$ produces a recombination of the originally intersecting
D-branes (see Fig.\ \ref{nofluxfig}).  It is easy to see that when one
diagonalizes the solution (\ref{sol}) by use of a gauge transformation
one obtains
\begin{eqnarray}
 Y = \left(
\begin{array}{cc}
\lambda(x) & 0 \\ 0 & -\lambda(x)
\end{array}
\right),
\qquad \lambda(x)\equiv \sqrt{p^2x^2 + a^2} \ .
\end{eqnarray}
Thus, it seems that the two D-strings are disconnected for any nonzero
value of $a$.  We show in the following subsection by analyzing the
D-string charge of the configuration that this intuitive picture is
correct: the D-strings are indeed localized on the curves pictured in
Figure.\ \ref{nofluxfig}.  In the configuration we are considering
here, however, the fundamental strings, originally bound in the
D-strings, must now stretch from one D-string world-volume to the
other, and thus live in the space ``between'' the branes.

\subsection{F-string charge}
\label{sec:charge}

Let us now proceed to study the properties of the supersymmetric
solution (\ref{sol}).  We are particularly interested in understanding
the structure of the F-strings running between the disconnected
D-strings located at $y=\pm \lambda(x)$.

The bulk supergravity currents arising from matrix configurations of
multiple D-branes were obtained in \cite{WT-Mark}.  We simply follow
the definition given there for the fundamental string current (see
also \cite{Myers,Okawa-Ooguri,lozano}).  The source current for the
NS-NS $B_{MN}$ field is given in our case by
\begin{eqnarray}
 \widetilde{\Pi}_{10}(x,k)
&= &T_{\rm D1}(2\pi\alpha')
{\rm Str}\left[e^{ikY}F_{10}\right] 
\nn\\
 \widetilde{\Pi}_{20}(x,k)
&= &T_{\rm D1}(2\pi\alpha')
{\rm Str}\left[e^{ikY}F_{10}\partial_1Y\right] 
\label{currentform}
\end{eqnarray}
where $k$ is the momentum along the $y$ direction, and Str is the
symmetrized trace, where $Y, F$, and $\partial_1 Y$ are treated as
units in the symmetrization.\footnote{In most of the relevant
literature the gauge $A_0=0$ is employed, but here we have used a
gauge transformation to derive the currents in the $A_1 = 0$ gauge
used here.}  The symmetrized trace which appears in these
expressions plays a crucial role in the combinatorics of the current
analysis we carry out here and later in the paper.  The necessity of
using this particular resolution of the ordering ambiguities in the
current was found in matrix theory in \cite{Dan-WT,WT-Mark-2} and was
derived from string disk amplitudes in \cite{Okawa-Ooguri}.  In the
expressions (\ref{currentform}) the components $\tilde{\Pi}_{i0}$ are
Fourier transforms of a string charge density $\Pi^{0i} (x, y)$ which
couples to the $B$-field through $B_{0i}\Pi^{0i}$.  The $y$-dependence
of $\Pi^{0i}$ is defined implicitly through (\ref{currentform})
through the appearance of the matrix $Y$ in the symmetrized trace; for
calculational purposes, we can think of the Fourier transform
$\tilde{\Pi}$ as being defined as an infinite sum over multipole
moments of $\Pi^{0i}(x, y)$ in the $y$-direction.

We now proceed to evaluate $\Pi_{10}$. Using 
$Y^2 = (p^2 x^2 + a^2)\identity$,
the symmetrized trace is simplified
\begin{eqnarray}
 \widetilde{\Pi}_{10}/T_{\rm D1}& = & \sum_n \frac{(ik)^n}{n!}
{\rm tr}\left[Y^n p \sigma_3\right] 
\nn\\
&= & \sum_{n\ {\rm odd}}
\frac{(ik)^n}{n!} (p^2 x^2 + a^2)^{(n-1)/2}
{\rm tr}\left[Y p \sigma_3\right] 
\nn\\
&= & \frac{2p^2x}{\sqrt{p^2 x^2 + a^2}}\sum_{n\ {\rm odd}}
\frac{(ik)^n}{n!} \lambda(x)^{n/2}
\nn\\
& = & 
\lambda'(x)
\left(e^{ik\lambda(x)}-e^{-ik\lambda(x)}\right) \,.
\end{eqnarray}
Thus, after the Fourier transformation back to the coordinate
representation we have
\begin{eqnarray}
 \Pi_{10}(x,y)= -\int\! dk \; e^{-iky} \widetilde{\Pi}_{10}(x,k)
=
T_{\rm D1}\lambda'(x)
\bigl[\delta(y-\lambda(x)) - \delta (y+\lambda(x))\bigr] \,.
\label{eq:p1}
\end{eqnarray}
This shows that this component of the current is non-vanishing only on 
the location of the D-strings, $y=\pm\lambda(x)$, and can be thought
of as an $x$-component of a string charge density localized to the D-strings.

\begin{figure}[tbp]
\begin{center}
\begin{minipage}{13cm}
\begin{center}
\includegraphics[width=6cm]{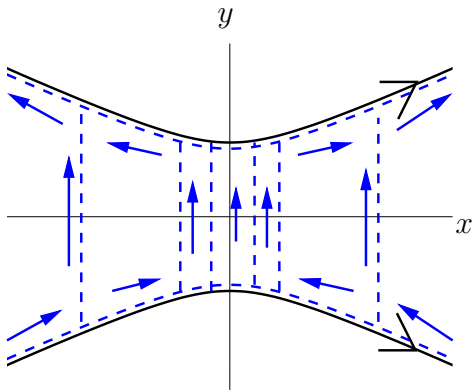}
\put(-90,140){$y$}
\put(0,60){$x$}
\caption{Arrows indicate the directions of the flux current for the
 fundamental strings. The dashed lines denote the current itself. Note
 that the vertical dashed lines are actually 
smeared along $x$ although they look
localized in this figure.}
\label{juncfig}
\end{center}
\end{minipage}
\end{center}
\end{figure}

The evaluation of the $y$-component of the current is a bit more
involved. Note that the symmetrized trace gives two cases
\begin{eqnarray}
 {\rm tr} \left[ Y F_{01} Y F_{01}\right]= 2p^2 (p^2 x^2 - a^2) \ , \\
 {\rm tr} \left[ Y^2 F_{01}^2 \right]= 2p^2 (p^2 x^2 + a^2) \ .
\nonumber
\end{eqnarray}
The precise combinatorial factors then give us
\begin{eqnarray}
 \frac{\widetilde{\Pi}_{20}(x,k)}{T_{\rm D1}}
&=&
\sum_{n\ {\rm even}} \frac{(ik)^n}{n!}{\rm Str} 
\left[Y^n F_{01}^2\right](2\pi\alpha')^2 \nn\\
&=&
\sum_{j=0}^{\infty} \frac{(ik)^{2j}}{(2j)!}
(p^2x^2+a^2)^{j-1}
{\rm tr} 
\left[\frac{j+1}{2j+1} Y^2F_{01}^2
+ \frac{j}{2j+1}YF_{01}Y F_{01} \right](2\pi\alpha')^2
\nn\\
&=&
2p^2 \sum_p \frac{(ik)^{2j}}{(2j)!}(p^2x^2+a^2)^j
-4a^2 p^2 \sum_p \frac{(ik)^{2j}}{(2j+1)!}j(p^2x^2+a^2)^{j-1}
\nn\\
&=&
 p^2 \left(e^{ik\lambda(x)}+ e^{-ik\lambda(x)}\right)
-4a^2 p^2 \frac{\p}{\p (a^2)}
\sum_j \frac{(ik)^{2j}}{(2j+1)!}(p^2x^2+a^2)^{j}
\nn\\
&=&
 p^2 \left(e^{ik\lambda(x)}+ e^{-ik\lambda(x)}\right)
-4a^2 p^2 \frac{\p}{\p (a^2)}
\left[\frac{1}{2ik\lambda(x)}
\left(e^{ik\lambda(x)}- e^{-ik\lambda(x)}\right)
\right]
\nn\\
&=&
 p^2 \left(1-\frac{a^2}{\lambda(x)^2}\right)
\left(e^{ik\lambda(x)}+ e^{-ik\lambda(x)}\right)
+\frac{a^2p^2}{ik\lambda(x)^3}
\left(e^{ik\lambda(x)}- e^{-ik\lambda(x)}\right)
\nn\\
&=&
 \lambda'(x)^2
\left(e^{ik\lambda(x)}+ e^{-ik\lambda(x)}\right)
+\frac1{ik} \lambda''(x)
\left(e^{ik\lambda(x)}- e^{-ik\lambda(x)}\right) \ .
\end{eqnarray}
After the Fourier transform, the current is 
\begin{eqnarray}
 \Pi_{20}(x,y)&= & T_{\rm D1}\lambda'(x)^2 \bigl[
\delta(y-\lambda(x)) + \delta(y+\lambda(x))
\bigr]\label{eq:2-current}\\
& &\hspace{0.6in}
+ T_{\rm D1}\lambda''(x) \bigl[
\theta(y+\lambda(x)) - \theta(y-\lambda(x))
\bigr] \ .\nonumber
\end{eqnarray}
Here $\theta$ is  the step function with
\begin{eqnarray}
 \theta(z)=
\left\{
\begin{array}{ll}
1 & (z>0) \\
0 & (z<0)
\end{array}
\right.
\end{eqnarray}
The first term in the current (\ref{eq:2-current}) is
localized on the D-strings as before.  The second term, on the other
hand, is non-vanishing in the region $-\lambda(x)< y < \lambda(x)$.
This can be interpreted as indicating that the string flux is flowing from
one D-string to another, vertically along $y$. The distribution of
string flux is sketched in Fig.\ \ref{juncfig}.  An interesting
feature of the $y$-component of the string current $\Pi_{20}$ is that
the integral of this flux over $y$ gives a quantity independent of $x$
\begin{eqnarray}
\int dy \; \Pi_{20} (x, y) & = &  2 (\lambda' (x))^2 + 2 \lambda (x)
 \lambda'' (x) = 2 \frac{d}{dx} (\lambda (x) \lambda' (x)) 
\label{eq:u}\\
 & = & 2p^2 \,. \nonumber
\end{eqnarray}

We have now calculated the full string current density $\Pi_{i0} (x,
y)$.  To check the consistency of this calculation, we can check that
this current is conserved,
\begin{eqnarray}
 \frac{\p}{\p x} \Pi_{10} +
 \frac{\p}{\p y} \Pi_{20} =0 \ .
\end{eqnarray}
This conservation rule follows directly from the preceding
expressions.

Note that while here we have not included higher-order $\alpha'$
corrections to the currents, we may expect that such corrections to
current conservation automatically cancel since the classical solution
of our concern is supersymmetric.  Indeed, for similar reasons we
expect that our solution also solves the full equations of motion in
the nonabelian generalization of Born-Infeld theory; this can easily
be checked explicitly to order $F^4$, and holds in 
some situations for the
symmetrized trace part of the nonabelian action \cite{HHS2}\footnote{
Note that in higher-dimensional configurations, however, the
supersymmetry conditions of the Yang-Mills theory may differ from those
of the Born-Infeld action \cite{jt}.
}
(related results in the abelian Born-Infeld theory are discussed in
\cite{Thorlacius}).  To see 
in slightly more detail how the currents are affected by higher-order
terms, consider for example the first leading correction to the
$x$-component of the string current \cite{WT-Mark}
\begin{eqnarray}
\widetilde{\Pi}_{10}(x,k)
=T_{\rm D1}(2\pi\alpha')
{\rm Str}\left[e^{ikY}F_{10}\left(1 
+ \frac12 (2\pi\alpha')^2 F_{10}^2 - \frac12 (\p_1 Y)^2\right) 
\right] \ .
\end{eqnarray}
In this expression, the higher-order terms in the parenthesis cancel
with each other. Although we have not checked all such higher-order
terms (there is no consistent proposal at this time for resolving
ordering ambiguities in the higher-order terms in the current), we
expect that supersymmetry protects both the solution we have found and
its currents.

To complete this subsection, we now also evaluate the D-string current
of our nonabelian brane configuration. 
The components of the D-string current are given in Fourier space by
\begin{eqnarray}
\widetilde{I}^{\rm D1}_{10} &= &T_{\rm D1} \mbox{Str} 
\left[e^{ikY}\right] \ ,  \label{D1current}
\\
\widetilde{I}^{\rm D1}_{20} &= & T_{\rm D1} \mbox{Str} \left[\p_1 Y
e^{ikY}\right] = \widetilde{\Pi}_{10} \ .
\nonumber
\end{eqnarray}
A similar computation to the above gives 
\begin{eqnarray}
I^{\rm D1}_{10} (x, y) &= &T_{\rm D1} 
\bigl[\delta(y-\lambda(x)) + \delta (y+\lambda(x))\bigr],
\\
I^{\rm D1}_{20} (x, y)&=  &T_{\rm D1} 
\lambda'(x)
\bigl[\delta(y-\lambda(x)) - \delta (y+\lambda(x))\bigr] \ .
\end{eqnarray}
Again, this current is conserved, and we observe that 
the D-string charge is located only on the curves $y=\pm\lambda(x)$.
This agrees with the eigenvalues of the field $Y$, as stated in the
previous subsection.

\subsection{Connection with string junctions}
\label{sec:junction}

\begin{figure}[bp]
\begin{center}
\begin{minipage}{11cm}
\begin{center}
\includegraphics[width=9cm]{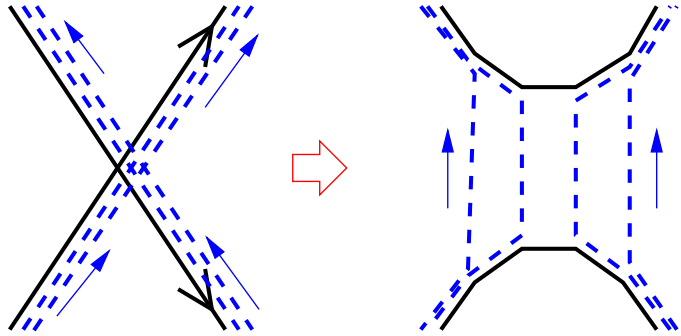}
\caption{Intersecting straight di-strings are deformed to 
 recombined di-strings connected by a large number of F-strings. 
 This deformation does not cost any energy and is thus an exactly marginal
zero mode of the configuration.}
\label{f:many-f}
\end{center}
\end{minipage}
\end{center}
\end{figure}

We now discuss how the configuration we have constructed is related to
the string junctions and networks studied in
\cite{das,ggt,Sen-network}.  In \cite{das}, BPS string junctions
between three $(p, q)$ di-strings are described as singular string
configurations.  In the world-volume theory, such a junction is
described by including singular source terms.  These singular
junctions were combined to form BPS networks of di-strings in
\cite{Sen-network}.  In our construction, we have a completely smooth
description of a classical brane system.  While the number of
D-strings in our system is quantized and finite (2), since we are
working in the classical Yang-Mills limit, F-string charge (electric
flux) is not quantized.  Thus, we should think of our configuration
as describing a limiting case of a string network with a large number
of fundamental strings connecting two D-strings (see
Figure~\ref{f:many-f}).  It is interesting to ask how in the quantum
theory the quantization of string flux will emerge.  Indeed, in the
full quantum Born-Infeld theory, we would expect to be able to
construct a class of configurations where the F-string and D-string
geometries are equivalent under S-duality (and an exchange of the
$x-y$ axes).  We discuss this question briefly at the end of this
section, and again in Section 4.

Let us now check quantitatively that the configuration we have
constructed can indeed be interpreted as a continuous limit of a
family of networks with many F-strings.  First,
consider the local flow of F-string charge along the D-string.  As we
can see from the asymptotic behaviour of $\Pi_{10}$, the lower
D-string loses $2p$ units of electric flux as $x$ goes from $-\infty$
to $+\infty$. This means that the fundamental strings stuck vertically
to the D-strings carry that quantity out from the lower D-string to
the upper D-string. This quantity can be evaluated by integrating the
second term of $\Pi_{20}$ over $x$,
\begin{eqnarray}
 \int \! dx \;  \lambda''(x) = 
\left[\lambda'(+\infty) - \lambda'(-\infty) \right]= 2p \ .
\end{eqnarray}
This relation  follows directly from current conservation,
but it also shows explicitly that at each point along the D-string
there is essentially a local junction which contributes a differential
part of
the total $2p$ units of F-string charge.  Since the whole
configuration is supersymmetric and therefore stable, the tensions of
the component strings should be locally balanced.  Let us now verify
that this is the case.

Since the vertical F-strings are smeared along the $x$ direction, 
the string junctions are correspondingly smeared. 
Let us concentrate on the distribution around the upper
D-string. The F-string current density can be divided into two vectors
as  
\begin{eqnarray}
\vec{{\bf \Pi}}\equiv 
 (\Pi_{10}, \Pi_{20} ) = 
\vec{{\bf \Pi}}_{\rm bound} \delta(y-\lambda(x))
+ \vec{{\bf \Pi}}_{\rm away} 
(\theta(y+\lambda(x))-\theta(y-\lambda(x)))\ , 
\end{eqnarray}
where $\vec{{\bf \Pi}}_{\rm bound}$ and $\vec{{\bf \Pi}}_{\rm away}$ 
are the F-string charge current bound in the upper D-string and away
from the D-strings, respectively :  
\begin{eqnarray}
\vec{{\bf \Pi}}_{\rm bound}
 \equiv T_{\rm D1}\lambda'(x) \,(1,\,\lambda'(x)) \ ,
\qquad 
\vec{{\bf \Pi}}_{\rm away} \equiv T_{\rm D1}\, (0,\,\lambda''(x)) \ .
\end{eqnarray}
Since $\vec{{\bf \Pi}}_{\rm bound}$ is parallel to
the upper di-string, the tension of this string at $x$ is given by
\begin{eqnarray}
 |\vec{\cal T}| = 
T_{{\rm D1}} \sqrt{1 + (\lambda' (x))^2} \,.
\end{eqnarray}
This indicates that the tension vector itself is 
simply given by
\begin{eqnarray}
 \vec{\cal T} = T_{\rm D1}(1, \lambda'(x)) \ .
\end{eqnarray}
The difference between the tension vectors at $x$ and $x + \Delta x$
is then given by 
\begin{eqnarray}
 \vec{\cal T}(x+\Delta x)- \vec{\cal T}(x) 
= \Delta x \, \vec{{\bf \Pi}}_{\rm away}  \ , 
\end{eqnarray}
which is just the tension of the vertical F-string times its density at
$x$  times the width $\Delta x$. This shows that the tensions are balanced
locally around the D-string.

Thus, we have seen that the configuration we have constructed is a
smooth limit of a family of string networks with F-strings vertically
connecting two D-strings.  From (\ref{eq:u}), we see that this
configuration is characterized by having a uniform distribution of net
string charge in the $y$-direction.  It would be interesting to
construct other classical limits of string networks.  For example, it
seems that it should be possible, at least in the full Born-Infeld
theory, to construct string networks, like the brane box
configuration, with macroscopic F-string charge localized at certain
places on the $x$-axis.  We do not know, however, how to realize such
a construction explicitly in the Yang-Mills theory, or if indeed such
a construction is possible.

\subsection{Other string networks}
\label{sec:generalization}

It is straightforward to generalize the preceding construction to a
very wide class of string networks.  Consider a generic string network
with the property that as $x \rightarrow \infty$ the set of asymptotic
$(p, q)$ strings is the same as that found when $x \rightarrow
-\infty$.  We expect that any such network can be constructed using
the same general approach used in the preceding discussion.  Indeed,
we can construct a general set of matrices
\begin{equation}
 Y = 2\pi\alpha'A_0 =  x \; {\rm Diag} (p_1, p_2, \ldots p_n) + C
\label{eq:network-form}
\end{equation}
satisfying (\ref{eq:bps}), where $p_i$ denotes the slope of the $i$th
D- string (which can be a fraction $p/q$ when $q$ of the $p_i$'s are
identical), 
$C_{ii}$ give the $y$-intercepts of the asymptotic $(p, q)$ strings,
and the off-diagonal matrix elements $C_{ij}$ parameterize  a moduli
space of BPS configurations with these asymptotic di-string charges.

\begin{figure}[bhtp]
\begin{center}
\begin{minipage}{14cm}
\begin{center}
\begin{minipage}{4.5cm}
\begin{center}
\includegraphics[width=4.5cm]{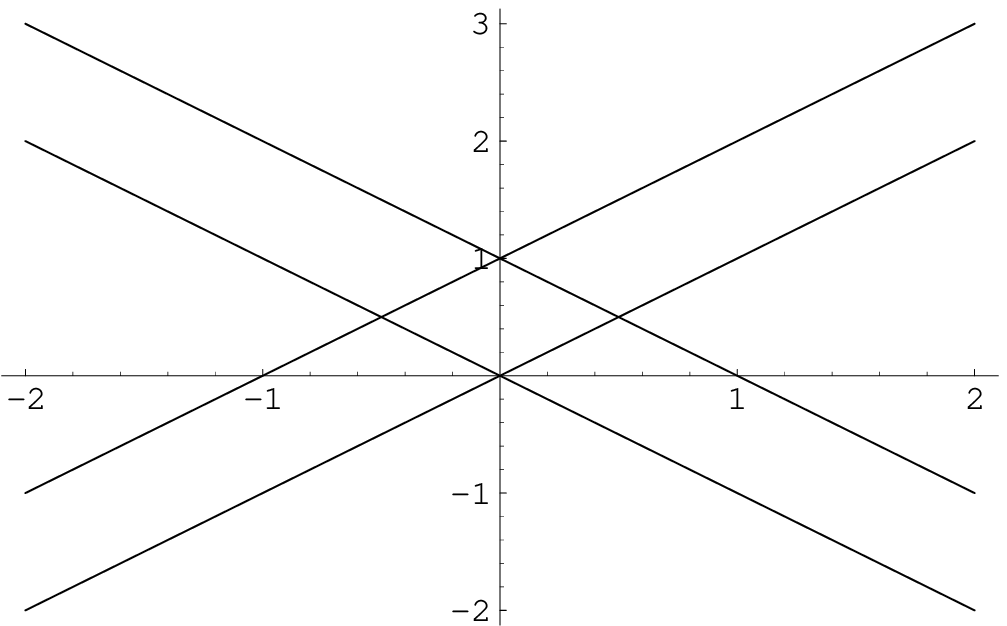}
(a) $a = 0$
\end{center}
\end{minipage}
\begin{minipage}{4.5cm}
\begin{center}
\includegraphics[width=4.5cm]{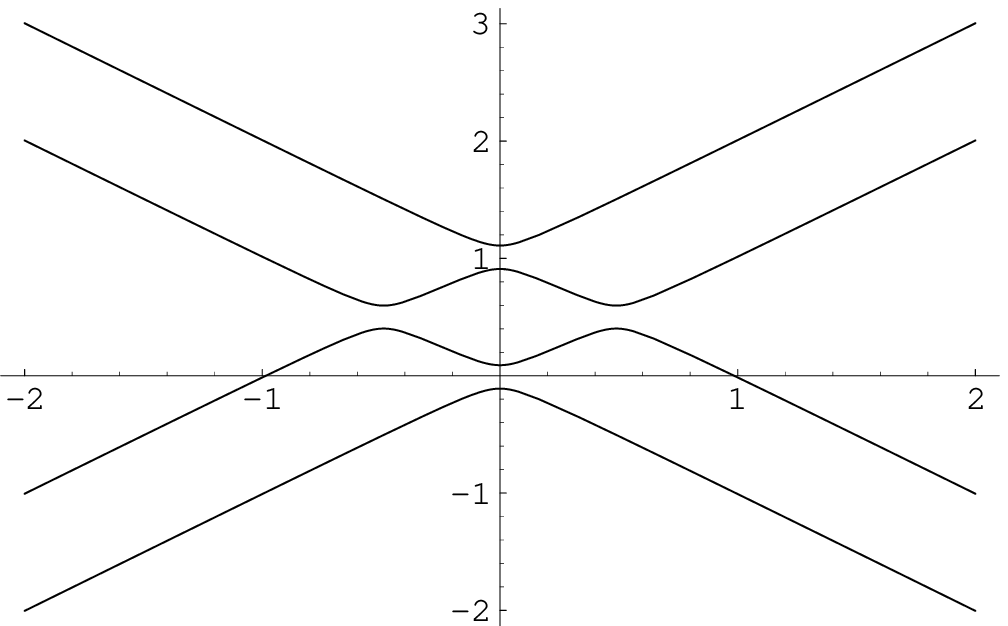}
(b) $a = 0.1$
\end{center}
\end{minipage}
\begin{minipage}{4.5cm}
\begin{center}
\includegraphics[width=4.5cm]{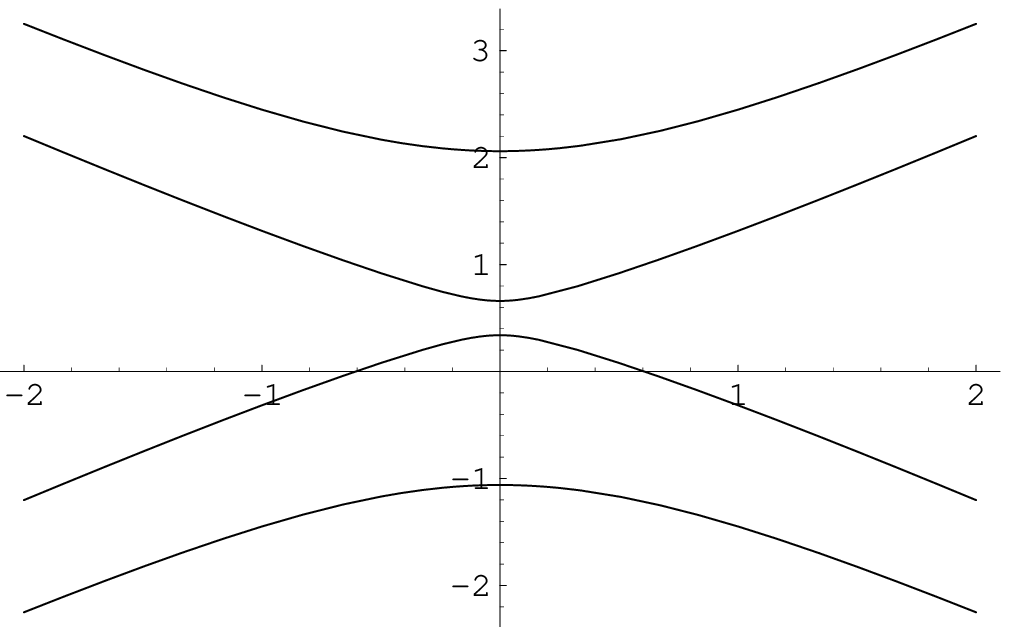}
(c) $a = 0.7$
\end{center}
\end{minipage}
\end{center}
\caption{String network of four intersecting branes with various
  values of the parameter $a$.}
\label{f:network-4}
\end{minipage}
\end{center}
\end{figure}

As an example of this string network construction, consider the
configuration 
\begin{equation}
 Y = 2\pi\alpha'A_0 =  
\left(
\begin{array}{cccc}
x & a & 0 & a\\
a & -x & a & 0\\
0 & a & x + 1 & a\\
a & 0 & a & -x + 1
\end{array}
\right) \ .
\end{equation}
This describes a network of
four asymptotic
di-strings with slopes
\begin{eqnarray*}
p_1 = p_3 & = &  1\\
p_2 = p_4 & = &  -1
\end{eqnarray*}
and $y$-intercepts
\begin{eqnarray*}
C_{11} = C_{22} & = &  0\\
C_{33} = C_{44} & = &  1 \ .
\end{eqnarray*}
All intersections between strings with opposite slopes are blown up by
the same parameter $a$.  Diagonalizing this matrix, we have
eigenvalues
\begin{equation}
Y_{\sigma \tau} =
\frac{1 + \sigma 
\sqrt{1 + 8a^2 + 4x^2 + 4 \tau  \sqrt{a^2 + 4a^4 + x^2}}}{2} 
\end{equation}
where $\sigma, \tau \in\{\pm 1\}$.  These eigenvalues are graphed for
various $a$ in Fig.~\ref{f:network-4}.

Just as for the simpler configuration with a single intersection
discussed in the previous subsections, the F-string flux for this and
other networks can be calculated and shown to give a smooth
distribution; as before, this distribution will include a piece which
is bound to the D-strings, and a piece which stretches between them,
corresponding to a smooth family of string junctions.

By using an infinite number of branes, we can construct a periodic
string network.  Considering the infinite periodic space as the
covering space of a torus, we can carry out T-duality on the periodic
network \cite{WT-T-duality}, giving us a T-dual configuration
consisting of D2-branes with momentum encoding the T-dual of the
electric flux (string
winding).  More precisely, consider an infinite brane configuration
described by matrices of the form (\ref{eq:network-form}), where $i
\in{\bf Z}$,
\begin{equation}
p_i = (-1)^i 
\end{equation}
and
\begin{equation}
C_{ij} =\left\{\begin{array}{l}
\lfloor\frac{i-1}{2} \rfloor R, \; \; i = j\\
a, \; \; i \neq j, i + j \equiv 1 \; ({\rm mod}\, 2)\\
0, \; \; i \neq j, i + j \equiv 0 \; ({\rm mod}\, 2)
\end{array} \right. 
\end{equation}
This encodes an infinite array of equally spaced di-strings with
slopes $\pm 1$, with equal parameters describing the deformation of
each intersection.  Performing a T-duality transformation in the $Y$
direction, we find a configuration of two D2-branes with gauge field
\begin{equation}
A_0 = A_y = \frac{1}{2\pi\alpha'}\left(\begin{array}{cc}
x & f(y)\\
f(y) & -x
\end{array}\right) \, 
\quad \mbox{where} \quad
f(y) \equiv 2\pi a R' \sum_n \delta(y-2\pi R'n) \ .
\end{equation}
Here $R'\equiv 2\pi\alpha'/R$ is the dual radius.

This construction gives us a BPS configuration of U(2) Yang-Mills
theory on a two-torus, which has two units of D2-brane charge and 2p
units of momentum (T-dual to F-string charge in the $y$-direction).
In the gauge we have used here, the gauge field becomes singular.  It
would be interesting to study this system further, particularly in the
quantum theory.  As for our previous examples, this configuration has
a momentum distribution which is uniform in the $x$-direction, but it
should in principle be possible to construct periodic configurations
analogous to the brane box in Figure~\ref{boxfig} which have
nonuniform momentum distributions.  It would be interesting to
understand such configurations better.

%%%%%%%%%%%%%%%%%%%%%%%%%%%%%%%%%%%%%%%%%%%%%%%%%%%%%%%%%%%%%%%%%%%%
%%%%%%%%%%%%%%%%%%%%%%%%%%%%%%%%%%%%%%%%%%%%%%%%%%%%%%%%%%%%%%%%%%%%
%%%%%%%%%%%%%%%%%%%%%%%%%%%%%%%%%%%%%%%%%%%%%%%%%%%%%%%%%%%%%%%%%%%%

\section{The 't Hooft-Polyakov monopole}

In the previous section we have constructed a family of BPS string
networks in classical U(N) Yang-Mills theory.  A striking feature of
these configurations is that they contain fundamental strings which
stretch through a region of space not contained in the D-string
world-volume, although the fields on the D-strings are used to
construct the Yang-Mills theory.

In this section we consider a much more familiar construction: the 't
Hooft-Polyakov monopole.  The Prasad-Sommerfield U(2) monopole
\cite{Prasad-Sommerfield} is a simple solution of U(2) Yang-Mills
theory in 3 + 1 dimensions with a scalar field.  In supersymmetric
Yang-Mills theory, this configuration is BPS.  In \cite{Aki}, a nice
geometric picture of this solution was given, wherein the scalar Higgs
field is interpreted as describing the shape of a pair of D3-branes
connected by a ``tube'' which shrinks to a point and reverses
orientation halfway between the D3-branes.  The monopole solution has
magnetic flux, corresponding to D-string charge on the D3-branes.  We
use the nonabelian current analysis method to study where this
D-string current lives, and we show that just as for the string
networks of the previous section, part of the D-string flux lives on
the D3-branes, while another part passes through the space between the
branes.

\subsection{The BIon}

\begin{figure}[htp]
\begin{center}
\begin{minipage}{10cm}
\begin{center}
\includegraphics[width=6cm]{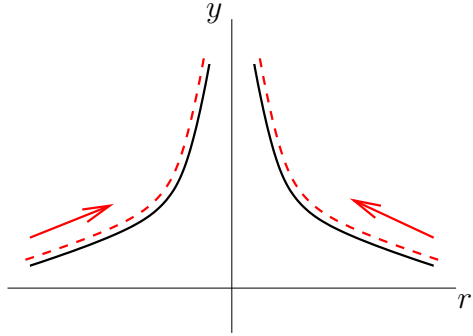}
\put(-95,120){$y$}
\put(0,10){$r$}
\caption{A slice of the BIon. The dashed lines with arrows are the
 D-string charges bound on the D3-brane spike (solid lines).}
\label{bionfig}
\end{center}
\end{minipage}
\end{center}
\end{figure}

Before going to the 't Hooft-Polyakov monopole, let us first consider a
BIon solution corresponding to a BPS abelian monopole; because this
configuration exists in the abelian theory, it is much simpler. The
BIon solution of the abelian $3+1$-dimensional Dirac-Born-Infeld
action is \cite{CM, Gibbons}
\begin{eqnarray}
 B^i =\frac{1}{2} \epsilon^{ijk}F_{jk}
=\frac{-bx^i}{ r^3}, \quad \Phi = \frac{b}{r} \
, \label{eq:abelian-monopole} 
\end{eqnarray}
which solves the BPS equation $B_i=\p_i \Phi$.
Here $r \equiv \sqrt{(x_1)^2 + (x_2)^2 + (x_3)^2}$, $i=1,2,3$, 
and $b$ is a constant which is quantized as $b=n/2$ where $n$ is an
integer \cite{CM}.
$\Phi$ is a transverse scalar which we interpret as a fourth spatial
dimension  ($y$), with the rescaling $Y=2\pi\alpha'\Phi$.
The solution (\ref{eq:abelian-monopole})
is interpreted as a semi-infinite D-string attached perpendicularly to
a D3-brane \cite{CM,Gibbons}. 
Let us compute the D-string charge density for this configuration
by use of the current density formula for the D3-brane. 
In this case, the general nonabelian formula simplifies as all
commutators are dropped, and the D-string charge density is associated
with U(1) magnetic flux.  We have
\begin{eqnarray}
 \widetilde{j}^{0y}(k)= 
T_{\rm D3}(2\pi\alpha')^2
 B_i \p_i\Phi e^{ik
(2\pi\alpha')
\Phi} 
=T_{\rm D3}(2\pi\alpha')^2 \frac{b^2}{r^4} e^{ik(2\pi \alpha')b/r} \ .
\end{eqnarray}
So after a Fourier transformation, we obtain
\begin{eqnarray}
 j^{0y} = 
T_{\rm D3}
(2\pi\alpha')^2
\frac{b^2}{r^4}\delta(y-2
\pi\alpha'
b/r) \ .
\end{eqnarray}
In the same way, we obtain
\begin{eqnarray}
 j^{0i}= 
-T_{\rm D3}
(2\pi\alpha')
 \frac{bx_i}{r^3}\delta(y- 2
\pi\alpha'
b/r) \ .
\end{eqnarray}
This shows that the D1-brane current lies completely on the deformed
D3-brane surface located at $y=2\pi\alpha' b/r$ (see
Fig.~\ref{bionfig}).  
Furthermore, it is easy to see that the D-string current is tangent to
the D3-branes, and has total flux which integrates at any $y$ to
\begin{eqnarray}
\int\!d^3x\; j^{0y}=
4\pi(2\pi\alpha')T_{\rm D3} b = n T_{\rm D1}.
\end{eqnarray}
Thus, in this abelian case  
there is no D-string charge ``away from'' the D3-brane, and there are
$n$ units of D-string flux going through a sphere at any finite radius
on the D3-brane.

\subsection{The 't Hooft-Polyakov monopole}

\begin{figure}[htp]
\begin{center}
\begin{minipage}{10cm}
\begin{center}
\includegraphics[width=8cm]{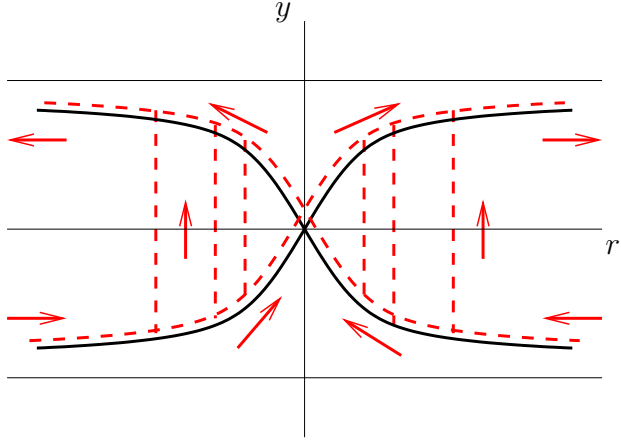}
\put(-125,160){$y$}
\put(0,70){$r$}
\caption{D-string current distribution (dashed lines) of the 't
 Hooft-Polyakov  monopole. }
\label{f:monopole}
\end{center}
\end{minipage}
\end{center}
\end{figure}

Let us now consider the 't Hooft-Polyakov monopole.  The BPS
equation  
\begin{eqnarray}
 \frac12 \epsilon_{ijk}F_{jk} = D_i \Phi 
\label{bpseq}
\end{eqnarray}
admits the Prasad-Sommerfield solution 
\cite{Prasad-Sommerfield}
\begin{eqnarray*}
A_i^{a} & = &  \epsilon_{aij} (1-K (r)) \frac{x_j}{r^2} \,, \nonumber\\
A^a_0 & = &  0 \,,\\
\Phi^a & =& -H (r) \frac{x_a}{r^2}  \,,\nonumber
\end{eqnarray*}
where we have used the SU(2) decomposition
$A_\mu = \sum_{a = 1}^{3}  A_{\mu}^a (\sigma_a/2)$,
and where
\begin{eqnarray*}
K (r) & = & Pr/\sinh (Pr) \ , \\
H (r) & = &  Pr\coth (Pr) -1 \ ,
\end{eqnarray*}
for a constant $P$.
The Higgs field $\Phi$ can be diagonalized by a gauge transformation as
\begin{equation}
\Phi = \left(\begin{array}{cc}
 P\coth(Pr) -\frac{1}{r}  & 0\\
0 & -P\coth(Pr) +\frac{1}{r}
\end{array}\right) \,.
\end{equation}
After the proper rescaling $Y=2\pi\alpha' \Phi$, 
the location of the D3-brane is given by 
\begin{eqnarray}
 y= \pm \lambda(r) \equiv \pm \pi\alpha' \frac{H(r)}{r} \, .
\end{eqnarray}
This corresponds to a simple D3-brane geometry, shown in the solid
line of Figure~\ref{f:monopole}.  

Now let us consider the D-string
flux in this configuration.  Before explicitly computing this flux, we
can note that since the monopole solution is smooth at $r = 0$ we
cannot have the full flux living on the brane, since this would entail
a finite flux passing through the vanishing ``neck'' of the tube
connecting the branes.  Thus, we see qualitatively that some part of
the D-string flux must move off the D3-brane world-volume.
More explicitly, we have
\begin{eqnarray}
D_i \Phi = \frac12 \epsilon_{ijk}F_{jk} = 
-\frac{x_i}{r}\left(
\p_r \left(\frac{H}{r^2}\right)+ \frac{H(1-K)}{r^3}
\right)\frac12 x_a \sigma_a \, - \, \frac{HK}{r^2}\frac12 \sigma_i \ .
\end{eqnarray}
The components of the D-string current density are given by
\begin{eqnarray}
\widetilde{j}^{0y} & = & (2\pi\alpha')^2 T_{\rm D3}
\mbox{Str}
\left[
\frac{1}{2}(\epsilon_{abc}F_{ab}D_c\Phi)e^{ik
(2\pi\alpha')
\Phi}
\right] \ , \label{eq:D-string-current}
\\
\widetilde{j}^{0a} & = & 
2\pi\alpha' T_{\rm D3}
\mbox
{Str}
\left[\frac12 \epsilon_{abc}F_{bc}e^{ik
(2\pi\alpha')
\Phi}
\right] \ .\nonumber
\end{eqnarray}
A straightforward calculation similar to the previous section 
gives the following result
\begin{eqnarray}
 j^{0y} & = &  (2\pi\alpha')^2 T_{\rm D3}
\left[
\frac{1}{4r^4}(1\!-\!K^2)^2 \left(\delta(y-\lambda) + \delta (y+\lambda)
\right)
\right.
\label{eq0y}
\\
& &\hspace{60mm} \left.
+ \frac{HK^2}{2\pi\alpha' r^3} \left(
\theta(y+\lambda) -\theta (y-\lambda)
\right)
\right] \ ,
\nonumber
\\
 j^{0a} & = &  (2\pi\alpha') T_{\rm D3}\frac{x_a}{2r^3}
(1-K^2)
\left(\delta(y-\lambda) - \delta (y+\lambda)\right) \ .
\label{eq0a}
\end{eqnarray}
It is easy to check that this current density satisfies the current
conservation law, 
\begin{eqnarray}
 \p_y j^{0y}+ \p_a j^{0a}=0 \ .
\label{eq:cc}
\end{eqnarray}
The meaning of the terms in the current density expression (\ref{eq0y})
and (\ref{eq0a}) is obvious, analogous to the previous section: 
the current (\ref{eq0a}) and the first term in (\ref{eq0y}) describes
the D-string charge density bound on the D3-brane world-volume surfaces,
while the second term in (\ref{eq0y}) describes  straight D-strings
connecting the upper and lower D3-branes, away from the D3-brane
surface. This feature is analogous to the F-string current density
observed in the previous section. 

For large $r$, these current expressions coincide with the BIon case,
and one can see that the second term in (\ref{eq0y}) damps
exponentially. On the other hand, for small $r$, the first term in
(\ref{eq0y}) converges to a constant while the second term apparently
diverges, as
\begin{eqnarray}
 j^{0y} \sim (2\pi\alpha')^2 T_{\rm D3} 
\left[\frac{P^4}{36}
\left(\delta(y-\lambda) + \delta (y+\lambda)
\right)
+ \frac{P^2}{ 6\pi\alpha' r} 
\left( \theta(y+\lambda) -\theta (y-\lambda)\right)
\right] \ .
\end{eqnarray}
This is  still consistent with the comment above regarding the lack of a
singularity in $F$ at the
``neck'' $r=0$.  The second term in $j^{0y}$ goes
as $1/r$.  Although this component of the current is unbounded as $r
\rightarrow 0$, however,
since the distance between the branes goes as $r$ in
this limit, all multipole moments of the current are finite.

We have now seen that the shape of the D3-branes in the monopole
configuration must be affected by the D-strings stretching vertically
between the branes.  As in Section \ref{sec2}, it should be possible to
verify that the tension of the D-strings precisely produces the correct
amount of curvature of the D3-brane world-volume in the monopole
configuration. 
Indeed, based on a qualitative argument of this kind, the existence of
this D-string distribution away from the D3-brane surface was
conjectured in \cite{HHS2}. In Refs.~\cite{HHS1,HHS2,Kawano,LYi},
classical solutions of SU($N$) Yang-Mills theories corresponding to
string junctions ending on parallel D3-branes \cite{prong} (1/4 BPS
dyons) were constructed.  These constructions give natural
generalizations of the 't Hooft-Polyakov monopole and the Julia-Zee
dyon. In \cite{HHS2}, the shape of the deformed D3-brane surface in
this 1/4 BPS dyon was studied in detail, and it was observed that in
fact $(p,q)$-strings away from the deformed D3-brane surfaces are
necessary to consistently interpret the curved trajectories of the
D3-branes.  In this paper we have demonstrated that these strings are
actually existent, for the simplest case of the 't Hooft-Polyakov
monopole.  We expect that a similar computation of the NS-NS and R-R
current densities in the general 1/4 BPS dyon case will confirm the
more general conjecture in \cite{HHS2}.

%%%%%%%%%%%%%%%%%%%%%%%%%%%%%%%%%%%%%%%%%%%%%%%%%%%%%%%%%%%%%%%%%%%%
%%%%%%%%%%%%%%%%%%%%%%%%%%%%%%%%%%%%%%%%%%%%%%%%%%%%%%%%%%%%%%%%%%%%
%%%%%%%%%%%%%%%%%%%%%%%%%%%%%%%%%%%%%%%%%%%%%%%%%%%%%%%%%%%%%%%%%%%%

\section{Strings in the vacuum from Yang-Mills tachyons}
\label{sec3}

As an application of the methods developed in the previous part of
this paper, we now show how strings and D-strings in the vacuum appear
in Yang-Mills theory after tachyon condensation on a system of
intersecting branes.

Sen's conjectures on tachyon condensation \cite{conjectures} in
brane-antibrane systems have led to many interesting developments in
recent years.  In particular, these conjectures have led to a new wave
of development of string field theory, and to new efforts to
understand time-dependent processes in string theory.  In the context
of brane-antibrane annihilation in type II superstring theory, Sen's
conjectures essentially make three assertions:

\begin{itemize}
 \item[1.] Condensation of the tachyon  in a brane-antibrane system
describes the annihilation of the D-branes to the true vacuum.
\item[2.] All open string degrees of freedom are gone in the true
  vacuum, although closed strings and infinitely extended open strings
  still appear as degrees of freedom.
\item[3.]  Lower-dimensional D-branes appear as solitons of the
  tachyon field.  In particular,
vortex-like topological defects of the tachyon field
	  correspond to stable BPS D-branes with codimension two.
\end{itemize}

We will be interested here in parts 2 and 3 of these conjectures.  A
system containing a brane and an antibrane cannot be described in
Yang-Mills language, as there is no static gauge with respect to which
both the brane and antibrane have the same orientation.  It is,
however, possible to consider a pair of branes at an arbitrary angle
$\theta < \pi$ in Yang-Mills theory, just as we did for D-strings in
Section 2.  In the absence of supersymmetry-saturating fluxes, such a
pair of intersecting D-branes has a tachyon in the spectrum
\cite{bdl}.  The condensation of this tachyon is closely related to
the condensation of a brane-antibrane tachyon, and gives rise to a
dynamic recombination of the branes, leaving a true vacuum in a region
around the initial brane intersection locus.  This recombination can
be described in Yang-Mills language \cite{Aki-WT,Morosov,HN}, and has
been recently used in models of cosmology
\cite{intersection-cosmology}.  In this section we use the Yang-Mills
description of a pair of intersecting branes to give a qualitative
picture of how fundamental strings and codimension two D-branes appear
in the vacuum according to conjectures 2 and 3 above.

The question of how closed or infinitely extended fundamental strings
appear in the stable vacuum after tachyon condensation has sparked
quite a bit of recent work.  In particular,
it was argued in \cite{confinement,BHY,Sencon} that near the stable
vacuum an effective gauge theory for the brane system describes
fundamental strings as confined tubes of electric flux.
In this section we consider a pair of intersecting branes in the
presence of an electric flux.  We show that when the tachyon
condenses, the electric flux is converted into open strings stretching
through the vacuum between the recombined D-branes.  We use the
nonabelian current analysis methods from the previous sections to show
that the resulting strings indeed stretch between the branes.  This
gives a qualitative picture of how fundamental strings in the vacuum
appear in Yang-Mills theory. To study the third of Sen's conjectures,
we perform a similar analysis.  We turn on a vortex-shaped tachyonic
fluctuation of the Yang-Mills field on a D3-brane and compute the
D-brane currents in a similar way.  This shows how codimension two
D-branes stretch between the recombined D-branes.

In subsection \ref{sec:recombination-intro} we review the Yang-Mills
description of the recombination of intersecting branes through
tachyon condensation.  In subsection \ref{sec:strings-vacuum} we
describe how fundamental strings stretching between the recombined
branes arise from an initial electric field.  In subsection
\ref{sec3-3} we show how a vortex tachyon configuration gives rise to
D-strings stretching between recombined D3-branes.

\subsection{Sen's conjecture and recombination}
\label{sec:recombination-intro}

We begin by reviewing how the recombination process of a pair of
intersecting D-branes can be described in Yang-Mills theory.  This was
studied in a T-dual picture on the torus in \cite{Aki-WT} and for
branes in infinite flat space in \cite{Morosov,HN}.  Most of this
discussion follows the analysis of \cite{HN}.

Let us consider two D-strings for simplicity. The low-energy effective
description of parallel D-strings is given, as in Section 2, by the 1+1
dimensional SU(2) super Yang-Mills action (\ref{eq:l}).  We
decompose the matrix fields as $A_\mu = A^a_\mu (\sigma_a/2)$ where
$a=1,2,3$ and $\sigma_a$ are the sigma matrices.  The location of the
D-strings is specified by the eigenvalues of the transverse scalar
field $Y$, which is related to the usual Higgs field as
$2\pi\alpha'\Phi^a \equiv Y^a$.  The (unstable)
classical solution representing
the intersecting D-strings is
\begin{eqnarray}
 \Phi^3 = q x \ , \quad A_\mu = 0 \ ,
\label{backg}
\end{eqnarray}
where the linear coefficient $q$ is the slope of the D-strings, 
\begin{eqnarray}
 q = \frac{1}{\pi\alpha'}\tan(\theta/2) \; (>0)\ .
\label{relationq}
\end{eqnarray}
We have introduced the intersection angle $\theta$, which with
$\theta=0$ corresponds to a parallel pair of D-strings (which is BPS)
while $\theta=\pi$ represents an anti-parallel pair of D-strings
(parallel brane-antibrane).  The analysis of the spectrum of
fluctuations of this system shows that there are two tachyonic
fluctuation modes,
\begin{eqnarray}
 \Phi^1(x) = A^2_x(x) =  C^{(1)}(t)\exp 
\left[-\frac{qx^2}{2}\right], 
\;
 \Phi^2(x) = -A^1_x(x) =  C^{(2)}(t)\exp 
\left[-\frac{qx^2}{2}\right] \, , 
\label{gauss}
\end{eqnarray}
with the mass squared
$ m^2 = -q$ appearing in the free field equation for the fluctuation
modes, $(\p_0^2 -q) C^{(i)}(t)=0$.
For small $\theta$, 
this mass squared $m^2 = -q$ was found to be identical to that 
of the string worldsheet spectrum given in \cite{bdl}.

\begin{figure}[bhtp]
\begin{center}
\begin{minipage}{13cm}
\begin{center}
\includegraphics[width=9cm]{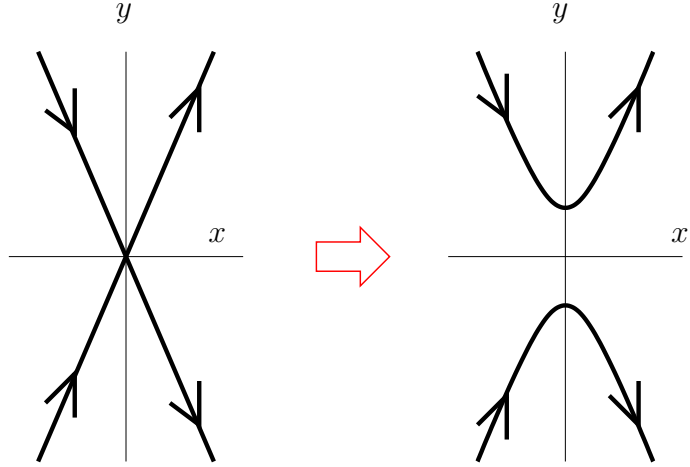}
\put(-50,170){$y$}
\put(-5,85){$x$}
\put(-215,170){$y$}
\put(-180,85){$x$}
\caption{D-strings are recombined. By rotating 
 these figures by $\pi/2$, one can see the local brane-antibrane
 annihilation occurring around the origin.}
\label{tachyonfig}
\end{center}
\end{minipage}
\end{center}
\end{figure}

Turning on the tachyon fields (\ref{gauss}) gives rise to a
recombination of the intersecting D-strings \cite{HN}.  In fact, when
only these modes are turned on, the eigenvalues of the field $Y$ are
given by
\begin{eqnarray}
 y=\pm \pi\alpha' \sqrt{q^2x^2+C^2 e^{-qx^2}} \ ,
\label{Ysol}
\end{eqnarray}
where $C^2 = (C^{(1)})^2 + (C^{(2)})^2$. The shape of the associated
D-strings is shown in Fig.~\ref{tachyonfig}.  Here we have
diagonalized the scalar field directly, but, as in Section 2, if we
plug the configuration with the tachyon turned on into the D-string
current formula (\ref{D1current}), we obtain the same result.

In this section of this paper we use Yang-Mills theory to study the
brane structure of field configurations analogous to (\ref{Ysol}) when
fundamental strings or codimension two D-branes are included in the
D-brane configuration through the addition of electric fields or a
topologically stabilized vortex configuration of the tachyon.  The
Yang-Mills action is only the leading part of the full brane action,
which is given by a nonabelian analogue of Born-Infeld theory (which
is not currently fully known).  The Yang-Mills description is accurate
at leading order in $\theta$, when the field strength $2 \pi \alpha'
F$ and $\theta$ are of comparable order and are both small compared to
1.  While we expect that the form of the tachyon modes (\ref{gauss})
will be essentially the same in Born-Infeld theory, the formula for
the mass should have $\tan (\theta/2) \rightarrow \theta/2$
\cite{bdl,Aki-WT,HN,naga}.  Even at finite $\theta$, however, we expect
that the Yang-Mills action should give a good qualitative description
of the physics of the tachyonic mode.

Thus, at a qualitative level, we can consider what happens as $\theta$
is of order unity.  When the angle $\theta$ is $\pi$, the intersecting
D-strings become a parallel D-string and anti-D-string.  At this
angle, we can no longer choose static gauge along the $x$-coordinate,
and we must instead parameterize the world-volume of the D-strings by
the static coordinate $y$.  At the point $\theta = \pi$, we cannot use
Yang-Mills theory to describe the system, but we expect that
qualitative features of brane-antibrane annihilation should be
captured in the Yang-Mills description of the theory in the limit
$\theta \rightarrow \pi$.  Let us view the intersecting D-strings (at
$\theta < \pi$) from the point of view of the D-brane world-sheet
parameter $y$.  As seen in Fig.\ \ref{tachyonfig}, when the tachyon
mode is turned on, a part of the ``world-volume'' parameterized along
the $y$ direction disappears due to the recombination, that is, the
local annihilation of the brane and the antibrane.  In this sense, the
Yang-Mills analysis we carry out here provides an interesting
realization of brane-antibrane annihilation.

At any angle $\theta$, the dynamical process of D-brane recombination
is very complicated.  A full treatment of this process would require
string field theory or the full nonabelian Born-Infeld action.  Even
in the simplified Yang-Mills action, this process is very complicated
since all string modes become involved after an infinitesimal time.
Nonetheless, when we consider very short time scales, only the tachyon
mode itself will be excited.  Thus, at short times, to understand the
geometry of the D-brane configuration just after recombination has
occurred, it will suffice to consider configurations where only the
tachyon field has been turned on, and where this field has a very
small coefficient $C$.  This is the spirit in which the following
analysis should be taken: the analysis is only technically accurate at
very small $\theta$ and at very small time.  Nonetheless, we believe
that it gives interesting insight into the way in which strings and
branes appear in the vacuum produced by tachyon condensation.

\subsection{Fundamental strings after tachyon condensation}
\label{sec:strings-vacuum}

\subsubsection{Correspondence to Yang-Mills theory}

In the usual language of the brane-antibrane near $\theta = \pi$, we
have two gauge fields: $A^{(+)}$ and $A^{(-)}$.  These are linear
combinations ($+$ and $-$) of two U(1) gauge fields living on the
brane and the antibrane.  When the tachyon condenses, the field
$A^{(-)}$ is Higgsed, since the tachyon is charged under the gauge
group relevant for this gauge field.  It has been conjectured
that the field $A^{(+)}$ is confined and gives rise to fundamental
strings when the original unstable brane-antibrane pair disappears
after the tachyon condenses \cite{confinement, BHY, Sencon}.

In the Yang-Mills description of intersecting D-strings, we also have
two U(1) gauge groups which are unbroken by the intersecting D-brane
solution.  This is consistent with the above observation; these two
descriptions are just related by changing the direction of the
world-volume parameterization.  In the Yang-Mills description, one
U(1) gauge field is the $\sigma_3$ component of the full SU(2)
gauge field.  The other U(1) is the overall trace
U(1), which decouples completely from the SU(2) sector.

To proceed, we have to check which U(1) in the Yang-Mills theory
corresponds to $A^{(+)}$.  One may naively think that the latter,
trace, U(1) should correspond to $A^{(+)}$, 
but this is not the case.
In changing the world-volume
parameterization from the horizontal direction $x$ (Yang-Mills
description) to the vertical direction $y$ (brane-antibrane
description), we must use the (leading-order) relation
\begin{eqnarray}
A_x^{\rm YM}=
\frac{\p Y}{\p x}
 A_y^{\rm brane-antibrane} \ .
\end{eqnarray}
Noting that the background (\ref{backg}) gives 
$\p Y / \p x \sim \sigma_3$, 
this relation shows that the Chan-Paton (CP) 
factor of the Yang-Mills field corresponding
to $A_y^{(+)}$ (whose CP factor is $\identity_{2\times 2}$) 
becomes in the brane-antibrane system
\begin{eqnarray}
  \sigma_3 =\sigma_3 \cdot \identity_{2\times 2} \ ,
\end{eqnarray}
while for $A_x^{(-)}$ (whose CP factor is $\sigma_3$), one has
\begin{eqnarray}
 \identity_{2\times 2} = \sigma_3 \cdot \sigma_3 \ .
\end{eqnarray}
In effect, the CP factors $\identity_{2\times 2}$ 
and $\sigma_3$ in the two
descriptions are exchanged,
and so we obtain the following correspondence: 
\begin{center}
\begin{tabular}{|c||c|c|}
\hline Brane-antibrane ($y$ direction)& $A^{(+)}$ & $A^{(-)}$ \\\hline
 Yang-Mills ($x$ direction)& $\sigma_3$ & $\identity_{2\times 2}$ \\\hline
\end{tabular}
\end{center}

\subsubsection{Fundamental strings and electric fields in $A^{(+)}$}

Let us first consider the field $A^{(+)}$. This field contains the
most interesting physics; it has been conjectured to be confined and
to provide fundamental strings after the tachyon has condensed.  From
the correspondence above, we know that the corresponding field in the
Yang-Mills theory is the $\sigma_3$ component of the SU(2) gauge
field.  In fact, we have already analyzed the effects of including an
electric field in this $\sigma_3$ sector in Section \ref{sec2}.
Although there we considered only the special case where the field
took the precise value needed to preserve the supersymmetries of the
system, the analysis of Section 2 already captures the intrinsic
features of the phenomena we are interested in.  

More precisely, let us consider modifying the intersecting brane
configuration (\ref{backg}) by including a small electric field
\begin{equation}
A_0 = \epsilon x \sigma_3 \,.
\end{equation}
As long as $\epsilon\ll q$, the inclusion of this field will not
change the dominant form of the tachyonic fluctuation modes
(\ref{gauss}).  Then, turning on the tachyon mode $C^{(1)}$
will give a
configuration which for small $x$ takes the form
\begin{equation}
 \Phi = \frac12 \left(
\begin{array}{cc}
q x  &  C \\  C & q x
\end{array}
\right)\, ,
\;\;\;\;\; 
A_0 = \left(
\begin{array}{cc}
\epsilon x  &  0 \\  0 & -\epsilon x
\end{array}
\right)\, , 
\;\;\;\;\; 
A_1 = \frac12\left(
\begin{array}{cc}
0  &  C \\  C & 0
\end{array}
\right)\,
.
\label{eq:tachyon-electric}
\end{equation}
For small $x$, we may discard the
off-diagonal components in $F_{10}$, which are proportional to $x$,
while the diagonal components are constant.
Since $F_{10}$ appears linearly in each component of the F-string
current density (\ref{currentform}), the F-string current in this
configuration is proportional to that computed in Section 2, and
is given by multiplying the current components (\ref{eq:p1},
\ref{eq:2-current}) by the overall factor $2\pi\alpha' \epsilon/p$,
where we take $p \rightarrow \pi\alpha' q$ and $a \rightarrow
\pi\alpha' C$.  This fundamental string distribution is depicted in
Figure~\ref{juncfig}.

Thus, as in the supersymmetric case discussed in Section 2, we
find that after tachyon condensation an electric flux on intersecting
branes is converted into fundamental strings which extend through the
region of space between the newly disconnected branes.
This gives a very satisfying picture in Yang-Mills theory of how Sen's
second conjecture is realized.  Even in the Yang-Mills approximation,
it seems, fundamental strings stretching through the vacuum can be
seen in the region of space where brane-antibrane annihilation has
occurred.  We expect that qualitatively similar results should hold in
the full Born-Infeld theory and in string field theory.

It should be reemphasized that this picture of strings in the vacuum
is fundamentally different than other pictures like the spike solution
of \cite{CM,Gibbons}, where F-strings stuck vertically onto a D-brane
world volume are realized as an electric flux on a cylindrically
deformed D$p$-brane world-volume whose spatial configuration has a
topology $R \times S^{p-1}$.  In that picture, the fundamental string
is taken to appear in the limit where the radius of the $S^{p-1}$
vanishes.  In the context of tachyon condensation, this realization of
the F-strings (or more precisely, a bound state of F-strings and 
D-branes) was recently used as a possible description of fundamental
strings after tachyon condensation \cite{HHW,HHNW,senbi}. On the other
hand, what we have found here is truly a representation of the F-strings 
``away from'' the D-branes.  The picture we have found here seems
to be much more natural from the point of view of Sen's conjectures,
where we expect the fundamental strings to exist in a region of
space-time completely devoid of D-brane matter.

One feature of the fundamental strings which is not as clear from this
picture, however, is the confinement mechanism of the strings.  As
discussed in Section 2, the classical configurations we have found
have F-string charge which is smeared in the $x$-direction.  
In the classical theory, fundamental string charge is
not quantized, so as for the supersymmetric configurations discussed
in Section 2, we expect that a quantum treatment will be needed to
fully understand the confinement and quantization of fundamental
strings in this Yang-Mills model.

\subsubsection{Electric field in $A^{(-)}$}

An electric flux for $A^{(-)}$ corresponds to a Yang-Mills background 
\begin{eqnarray}
 A_0 = \left(
\begin{array}{cc}
q' x & 0 \\
0 & q' x  
\end{array}
\right) \label{eq:em}
\end{eqnarray}
in addition to the original configuration (\ref{backg}), with some
constant $q'$.  This additional field is in the overall U(1)
sector, so there is no influence on the fluctuation analysis. This
means that there are still the same tachyonic fluctuation modes as in
(\ref{gauss}), and after the tachyon condensation the D-strings are
recombined.

\begin{figure}[htp]
\begin{center}
\begin{minipage}{13cm}
\begin{center}
\includegraphics[width=13cm]{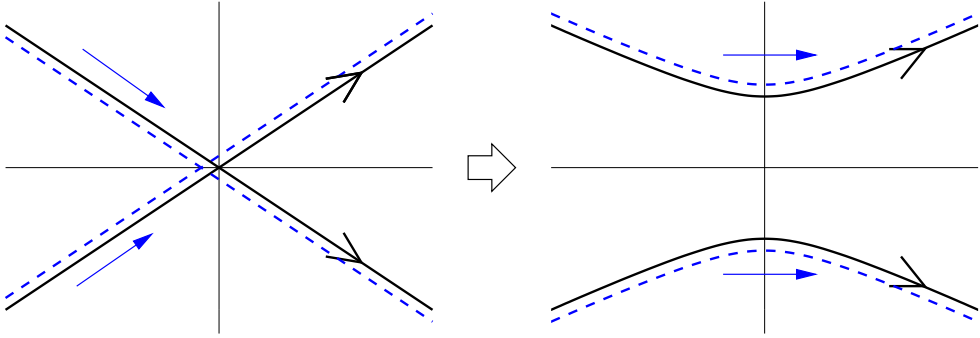}
\caption{Recombination of intersecting D-strings with $A^{(-)}$
 electric flux. The solid lines represent the D-strings, while the
 dashed lines represent F-strings. Note that the asymptotic
 orientations of the bound F-strings are different from those in section
 4.2.2, where the strings had to stretch between the
 separated branes. }
\label{ami}
\end{center}
\end{minipage}
\end{center}
\end{figure}

Let us see what happens to the F-string charge density. The charge
density formulae are given in (\ref{currentform}), and we substitute
the configuration (\ref{backg}, \ref{eq:em}) with tachyon condensation
$C\neq 0$ into (\ref{currentform}). Now, since our electric field is
in the overall U(1) sector, $F_{10}$ in the formula
(\ref{currentform}) does not contribute to the computation except as
an overall constant factor, and as a result the F-string charge
density (\ref{currentform}) is simply proportional to the D-string
charge density (\ref{D1current}). Since we know that the D-strings are
recombined, then this shows that the F-strings are also recombined,
and so there remains no F-string connecting the recombined D-strings
(see Fig.~\ref{ami}). This contrasts with the above results for the
gauge field $A^{(+)}$, and is consistent with the brane-antibrane
picture in which the gauge field $A^{(-)}$ is Higgsed after
tachyon condensation.

\subsection{Descent relation in Yang-Mills theory}
\label{sec3-3}

Let us now turn our attention to the third conjecture, describing the
creation of lower-dimensional D-branes as tachyonic topological
defects on unstable D-branes. In this subsection we consider the
formation of a vortex-like tachyon which gives a codimension two BPS
D-brane in the brane-antibrane system.  We will specialize in
particular to the formation of D-strings connecting intersecting
D3-branes which recombine through tachyon condensation.  We see again
that the charge of the lower dimensional D-brane lies ``off'' the
recombined D3-brane world-volume.  The resulting charge distribution
is quite similar to that of the 't Hooft-Polyakov monopole analyzed in
the previous section.

\subsubsection{Vortex formation}

To consider vortex formation in the tachyon system, we first must
check that the tachyonic modes (\ref{gauss}) found in the Yang-Mills
theory of intersecting D-branes actually have nontrivial topology in
their vacuum manifold.  In fact, a gauge transformation with a
transformation parameter proportional to $\sigma_3$ rotates one of
these fluctuations into the other, so the tachyon potential has a
U(1) symmetry and is written only in terms of the modulus of the
complex tachyon fluctuation, $|C^{(1)} + i C^{(2)}|$.  Hence, we
expect a vortex tachyon which is topologically stable. This is
consistent with the fact that the brane-antibrane setup has the same
structure and our intersecting brane setup is continuously related to
it.

\begin{figure}[bhtp]
\begin{center}
\begin{minipage}{14cm}
\begin{center}
\includegraphics[width=14cm]{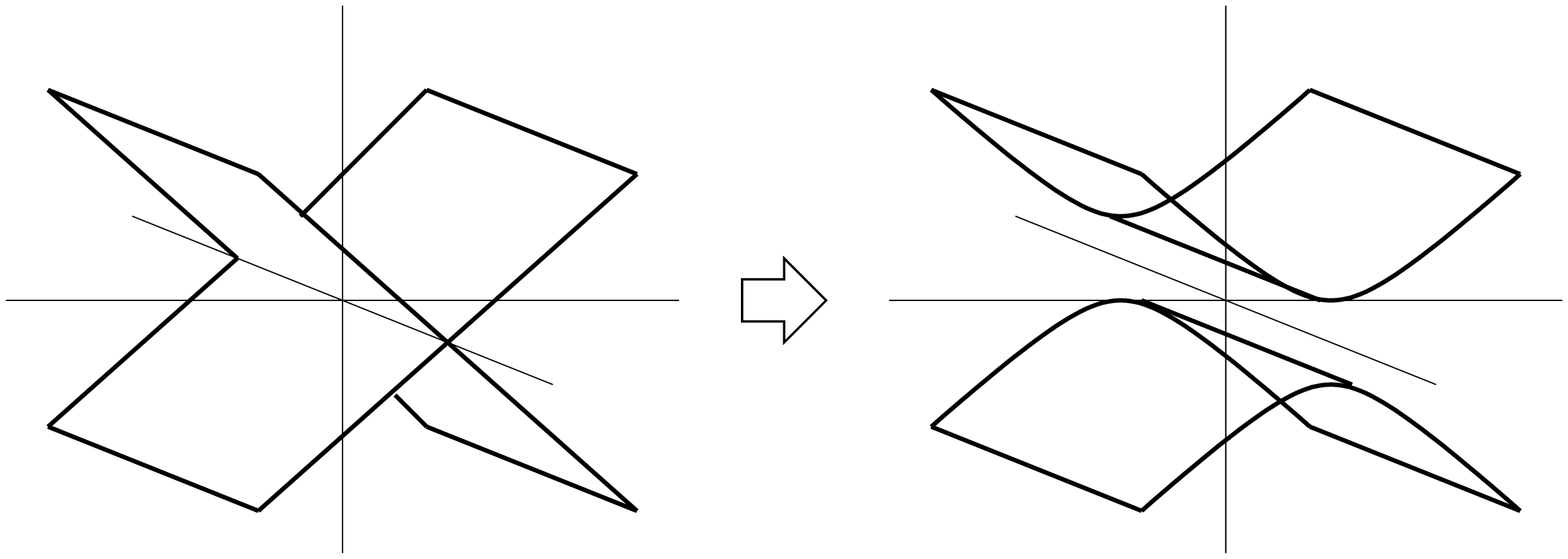}
\put(-100,140){$y$}
\put(-330,140){$y$}
\put(-230,70){$x_1$}
\put(0,70){$x_1$}
\put(-30,40){$x_2,x_3$}
\put(-250,40){$x_2,x_3$}
\caption{The intersecting D$3$-branes are recombined by the tachyon
 condensation which is homogeneous along the intersection submanifold.}
\label{d1fig}
\end{center}
\end{minipage}
\end{center}
\end{figure}

In this subsection we generalize the intersecting brane configuration
from D-strings to D3-branes so that we may have a tachyon vortex in
the $(2 + 1)$-dimensional intersection submanifold.  The D3-brane
world-volumes are parameterized by the coordinates $(t,x_1,x_2,x_3)$,
and the intersection submanifold is parameterized by $(t,x_2,x_3)$.
The coordinate $x_1$ plays the role of the world-volume coordinate $x$
in the intersecting D-string case.  Taking the coefficients of the
tachyon modes (\ref{gauss}) along $x_1$ to be functions of the
remaining coordinates $t, x_2, x_3$, we have the equation of motion
\begin{eqnarray}
[ (\p_t)^2 -(\p_2)^2-(\p_3)^2 + m_0^2] \;C^{(i)}(t,x_2, x_3)=0
\label{fluceq}
\end{eqnarray}
where negative $m_0^2=-q$ is the tachyon mass squared.\footnote{Of
course since we have more gauge fields on the world volume, the
fluctuation analysis should be examined again in this case.  It turns
out, however, that the tachyonic fluctuations still take the form
(\ref{gauss}) in the intersecting D3-brane case, although the profiles
of the higher massive modes are corrected.}  Since we have generalized
the configuration (\ref{backg}) trivially to the D3-brane
intersection, a homogeneous tachyon condensation (non-zero $C$)
generates the recombination shown in Fig.\ \ref{d1fig}.

Next we turn on a vortex-shaped tachyon,
\begin{eqnarray}
 C^{(1)} = cx_2 \ , \quad C^{(2)}=cx_3 \ ,
\label{linear}
\end{eqnarray}
where $c$ is a constant.\footnote{To be precise, this vortex-shaped
  tachyon represents  an unstable direction which will grow
  exponentially through the equation of motion.  If we solve the
fluctuation equation (\ref{fluceq}), we may obtain a time-independent
solution  for this unstable mode
\begin{eqnarray}
 C^{(1)} = (c/\sqrt{q})\sin \sqrt{q}x_2 \ , \quad 
 C^{(2)} =(c/\sqrt{q})\sin \sqrt{q}x_3 \ ,
\label{sinq}
\end{eqnarray}
where $c$ is a constant.  This represents a periodic array of
D-strings whose periodicity is tuned to give a marginal direction of
deformation.  We can expand this expression for small $x_2$ and $x_3$
and obtain the vortex configuration (\ref{linear}).  It is obvious
from this expression that the lower dimensional D-branes are created
as infinite pairs of brane-antibranes so that the total Ramond-Ramond
(RR) charge is preserved. In general, the functions $C^{(i)}$ are
time-dependent and describe dynamical creation and annihilation of the
lower dimensional D-branes.}  The linear profile (\ref{linear}) is
valid only for small $x_2$ and $x_3$.  As mentioned above, the
tachyonic fluctuations parameterize $S^1$ and so it is clear from the
asymptotic form that the mode (\ref{linear}) corresponds to the
formation of a tachyon vortex, which should give a codimension two
D-brane.

What we would now like to see is precisely how this Yang-Mills
configuration, together with the original background (\ref{backg}),
produces a lower-dimensional D-brane charge.  Let us consider what
kind of D-brane configuration is expected in the present case from
Sen's conjecture.  Firstly, at $\theta=\pi$ the brane configuration is
a parallel D3-brane-anti-D3-brane pair along $y$, so after the tachyon
condensation of the vortex in the $x_2$-$x_3$ world-volume space, the
topological defect is identified with a D-string which is orthogonal
to the $x_2$ and $x_3$ directions. This means that the D-string should
be oriented along the $y$ axis.  Secondly, we know from the preceding
discussion that the tachyon condensation on the intersecting D3-branes
realizes the usual D3-brane recombination.  Hence, combining these
predictions, we expect that condensation of the tachyon vortex
(\ref{linear}) will lead to a configuration containing a D-string
stretched between two recombined D3-branes.  We will now see that this
is indeed the case, by computing the location of the generated
D-string charges.  The D-strings are again found to live away from the
deformed D3-brane world-volume surface, just like the fundamental
strings discussed above and like the D-strings in the 't
Hooft-Polyakov monopole analyzed in the previous section.

\subsubsection{Shape of the D3-brane}

\begin{figure}[bhtp]
\begin{center}
\begin{minipage}{14cm}
\begin{center}
\includegraphics[width=14cm]{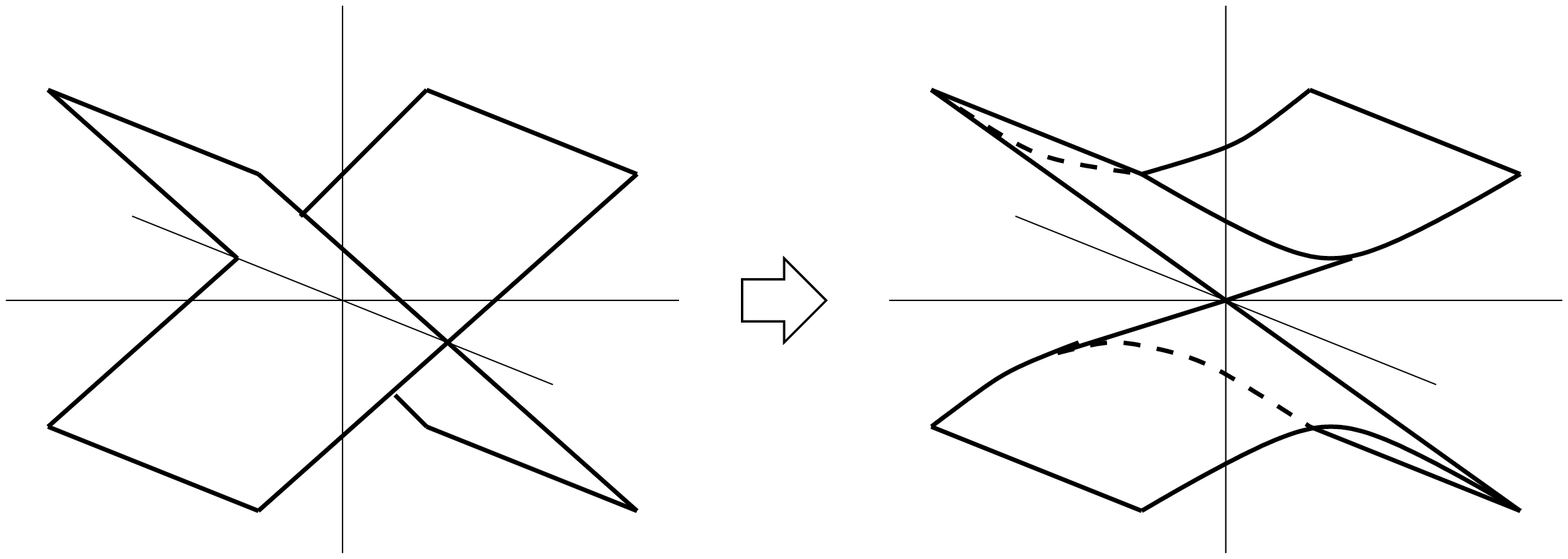}
\put(-100,140){$y$}
\put(-330,140){$y$}
\put(-230,70){$x_1$}
\put(0,70){$x_1$}
\put(-30,40){$x_2,x_3$}
\put(-250,40){$x_2,x_3$}
\caption{Plot of $y=\pm\lambda(x_1,x_2,x_3)$ after
intersecting D3-branes are recombined by a
vortex-like tachyon.}
\label{d1formfig}
\end{center}
\end{minipage}
\end{center}
\end{figure}

Once the vortex is given as (\ref{linear}), it is possible to study the
shape of the D3-brane world volume. The total scalar field configuration
is  
\begin{eqnarray}
 Y
= \pi\alpha'
(qx_1 \sigma_3 + cx_2 e^{-q(x_1)^2} \sigma_1+ cx_3 e^{-q(x_1)^2} \sigma_2) \ . 
\end{eqnarray}
The eigenvalues of this matrix, which give the location of the
D3-brane world-volume, are   then given by
\begin{eqnarray}
y= \pm \lambda(x_1,x_2,x_3) 
\equiv \pm 
\pi\alpha'
\sqrt{q^2 (x_1)^2 + c^2 e^{-2q(x_1)^2}((x_2)^2 + (x_3)^2)} \ ,
\end{eqnarray}
where $y$ is the bulk coordinate corresponding to the field $Y$.
This immediately shows that at the vortex core $x_2=x_3=0$ 
the intersecting
D3-branes still touch each other, while away from the origin those are
disconnected (recombined), as seen in Fig.~\ref{d1formfig}. This is
similar to the D3-brane deformation in the 't Hooft-Polyakov monopole
\cite{Aki} discussed in the previous section. 

\subsubsection{Generation of D1-brane charge}
\label{sec3-3-3}

In this subsection we explicitly compute the D-string RR charge
density produced by this Yang-Mills field configuration. From the RR
couplings in the D3-brane effective field theory, the relevant
components $j^{0 a}$ of the charge density are
again given by (\ref{eq:D-string-current}).
Let us first evaluate $j^{0y}$. 
Since again in this case $(\Phi)^2 = Y^2/ (2 \pi \alpha')^2$ is
proportional to the unit matrix,  
\begin{eqnarray}
 (\Phi)^2 = \frac14\left(q^2 (x_1)^2 
+ c^2 e^{-2q(x_1)^2}((x_2)^2 + (x_3)^2)\right)
\identity_{2\times 2}
=\frac{\lambda^2}{(2\pi\alpha')^2}\identity_{2\times 2} \ ,
\end{eqnarray}
the computation of the density current can be performed in the by-now
familiar fashion.  The field strengths are given by
\begin{eqnarray}
F_{23}=0, \quad F_{12}= D_3 \Phi= ce^{-q(x_1)^2}\sigma_2/2,\quad
F_{31}= D_2 \Phi=ce^{-q(x_1)^2}\sigma_1/2 \ .
\end{eqnarray}
Then using the results
\begin{eqnarray}
\mbox{tr}\left[\sigma_1 \Phi \sigma_1 \Phi\right]
&=& (-q^2 (x_1)^2 + c^2 e^{-2q(x_1)^2}((x_2)^2-(x_3)^2))/2 \ ,\\
\mbox{tr}\left[\sigma_1 \sigma_1 \Phi \Phi\right]
&=& (q^2 (x_1)^2 + c^2 e^{-2q(x_1)^2}((x_2)^2+(x_3)^2))/2 \ ,\\
\mbox{tr}\left[\sigma_2 \Phi \sigma_2 \Phi\right]
&=& (-q^2 (x_1)^2 + c^2 e^{-2q(x_1)^2}(-(x_2)^2+(x_3)^2))/2 \ ,\\
\mbox{tr}\left[\sigma_2 \sigma_2 \Phi \Phi\right]
&=& (q^2 (x_1)^2 + c^2 e^{-2q(x_1)^2}((x_2)^2+(x_3)^2))/2 \ ,
\end{eqnarray}
we obtain
\begin{eqnarray}
 \widetilde{j}^{0y}& = &T_{\rm D3} (2\pi\alpha')^2
\mbox{Str}\left[\sigma_1 \sigma_1 e^{ik(2\pi\alpha')\Phi}
+\sigma_2 \sigma_2 e^{ik(2\pi\alpha')\Phi}\right]c^2 e^{-2q(x_1)^2}\nn\\
& = &
T_{\rm D3}(\pi\alpha')^2
c^2 e^{-2q(x_1)^2}
\left[
\frac{\lambda^2-(\pi\alpha')^2q^2(x_1)^2}{\lambda^2}
(e^{ik\lambda}+e^{-ik\lambda})
\right.
\nn\\
& &
\left.
\hspace{50mm}
+
\frac{\lambda^2+(\pi\alpha')^2q^2(x_1)^2}{ik\lambda^3}
(e^{ik\lambda}-e^{-ik\lambda})
\right] \ .
\end{eqnarray}
After a Fourier transformation, we have
\begin{eqnarray}
 j^{0y} = 
T_{\rm D3}(\pi\alpha')^2c^2 e^{-2q(x_1)^2}
\left[
\frac{\lambda^2-(\pi\alpha')^2q^2(x_1)^2}{\lambda^2}
(\delta(y-\lambda)+ \delta(y+\lambda))
\right.
\hspace{10mm}
\nn\\
\left.
+
\frac{\lambda^2+(\pi\alpha')^2q^2(x_1)^2}{\lambda^3}
(\theta(y+\lambda)+ \theta(y-\lambda))
\right] \ .
\end{eqnarray}
This D-string charge density exhibits a by-now familiar structure.
\begin{figure}[htp]
\begin{center}
\begin{minipage}{13cm}
\begin{center}
\includegraphics[width=7cm]{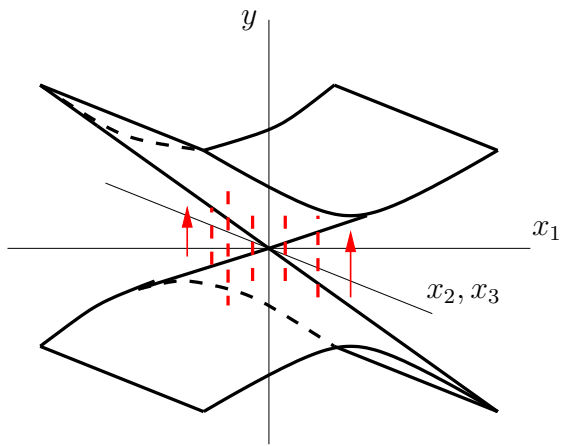}
\put(-110,160){$y$}
\put(0,80){$x_1$}
\put(-40,56){$x_2,x_3$}
\caption{A schematic view of the distribution of the D-string charge
 density (dashed lines) after tachyon condensation. Here we describe
 only the part of the D-string current which is away from the D3-brane
 surface (solid surface). The arrows denote the directions of the
 D-string charge density. }
\label{d1fluxfig}
\end{center}
\end{minipage}
\end{center}
\end{figure}
\begin{figure}[htp]
\begin{center}
\begin{minipage}{13cm}
\begin{center}
\includegraphics[width=6cm]{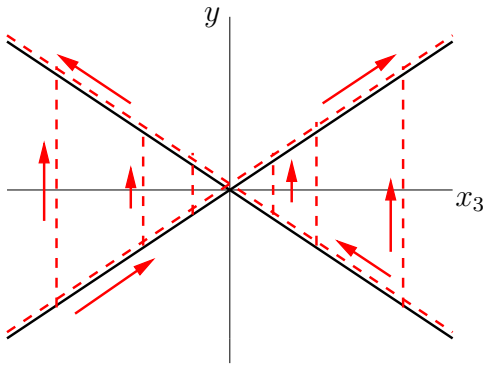}
\put(-95,130){$y$}
\put(0,60){$x_3$}
\caption{
 The distribution of the D-string charge density (dashed lines) in the
 slice $x_1=x_2=0$. Here again, the vertical dashed lines should be
 understood as the ones which are smeared along the horizontal
 direction. The arrows denote the directions of the D-strings. Solid
 lines denote slices of the D3-branes.} 
\label{d1slicefig}
\end{center}
\end{minipage}
\end{center}
\end{figure}
The first term describes D-strings running along the deformed D3-brane
surface $y=\pm\lambda(x_1,x_2,x_3)$.  The second term describes
vertical D-strings connecting the upper and the lower D3-branes. 
These D-strings lie off the world-volume of the D3-branes, and are
closely analogous to the D-strings found in the 't Hooft-Polyakov
monopole in the previous section.
It is easy
to evaluate the other components of the D-string current as
\begin{eqnarray}
j^{01} &=& 0 \ , 
\nn\\
j^{02} &=& \pi\alpha'T_{\rm D3}
x_2\frac{c^2 e^{-2q(x_1)^2}}{\lambda} \left(
\delta(y-\lambda)- \delta(y+\lambda)
\right) \ ,
\label{othercom}
\\
j^{03} &=& \pi\alpha'T_{\rm D3}
x_3\frac{c^2 e^{-2q(x_1)^2}}{\lambda} \left(
\delta(y-\lambda)- \delta(y+\lambda)
\right) \ .
\nn
\end{eqnarray}
One can verify that the current conservation law (\ref{eq:cc}) is
satisfied as a consistency check.\footnote{This current conservation
holds even though the configuration we are considering is not static,
since the current $j^{\mu\nu}$ is antisymmetric in its indices and the
conservation law for the charge density $\mu=0$ does not involve any
time derivative.}  As expected, the components (\ref{othercom}) of
the D-string current do not describe any D-strings located away from
the D3-branes,  but give only D-string charges along the D3-brane
surface. Fig.\ \ref{d1fluxfig} shows schematically how the D-string
charges are distributed in the bulk.  In Figure\ \ref{d1slicefig}, the D-string
charge density is shown in the spatial slice $x_1=x_2=0$.

We have thus shown that D1-branes suspended between recombined
D3-branes appear as a result of the formation of a vortex-like tachyon
on an intersecting pair of D3-branes.  This gives a simple Yang-Mills
description of how Sen's third conjecture on the formation of
lower-dimensional D-branes through topological tachyon defects is
realized in the case of the intersecting D-branes.  In this situation,
as in the description of fundamental strings in the previous
subsection, several caveats apply.  In particular, the analysis we
have carried out here is only valid for short times and small values
of $x$ near the core of the vortex.  Note, however, that unlike in the
case of fundamental strings, the quantization of D-string charge is
already seen in the classical picture presented here, as this
quantization arises simply from the quantization of the first Chern
class in the nonabelian gauge theory.

%%%%%%%%%%%%%%%%%%%%%%%%%%%%%%%%%%%%%%%%%%%%%%%%%%%%%%%%%%%%%%%%%%%%
%%%%%%%%%%%%%%%%%%%%%%%%%%%%%%%%%%%%%%%%%%%%%%%%%%%%%%%%%%%%%%%%%%%%
%%%%%%%%%%%%%%%%%%%%%%%%%%%%%%%%%%%%%%%%%%%%%%%%%%%%%%%%%%%%%%%%%%%%

\section{Conclusions and discussion}

In this paper we have analyzed a number of nonabelian D-brane
configurations containing fundamental strings and lower-dimensional
D-branes.  We used the nonabelian formulae for the fundamental string
and D-brane currents from \cite{WT-Mark} to compute the spatial
distribution of string and D-brane charges.  We found that in a
variety of situations, these charges extend through regions of space
away from the D-branes on which the original nonabelian theory was
defined.  This phenomena was demonstrated for string networks in
(1+1)-dimensional nonabelian Yang-Mills theory, for the 't
Hooft-Polyakov monopole in (3+1)-dimensional Yang-Mills theory, and
for tachyonic intersecting brane configurations, where we found
F-strings and D-strings in the vacuum formed in the brane annihilation
process.  The constructions we have presented here give an interesting
new way in which D-brane field theory encodes the geometry of
transverse directions of space-time.  

In Section 2 we constructed a new class of supersymmetric solutions of
the nonabelian Yang-Mills equations of motion.  These solutions
correspond geometrically to D-strings with fundamental strings
stretching between them, and give a smooth description in Yang-Mills
theory of string networks.  By performing the nonabelian analysis of
the current density, we showed that the fundamental string charge
density has support in regions of space-time away from the
world-volume of the D$1$-branes on which the Yang-Mills degrees of
freedom live.  There are a number of interesting open questions
related to these Yang-Mills string networks.  In particular, we do not
know how to construct networks where the total F-string charge in the
vertical (Higgs) direction is inhomogeneous in the spatial Yang-Mills
coordinate, or even whether such configurations can be constructed.
The Ansatz we used to solve the BPS equation (\ref{eq:bps}) is fairly
restrictive; it is possible that a more general family of solutions
may be found by loosening the constraints imposed by this Ansatz.

All the analyses we performed here were restricted to the Yang-Mills
approximation.  This is a good approximation for systems of D-branes
when the gauge field and transverse fluctuations are small, and gives
a qualitative picture of many important aspects of the physics of
D-branes.  In particular, we described particular limits in which the
picture of fundamental strings and D-strings in the vacuum arising
from tachyon condensation is valid.  For a full understanding of
string physics, however, it is necessary to extend the analysis to a
more complete formalism such as string field theory or nonabelian
Born-Infeld theory (which, when derivative corrections are included,
should describe all the physics of string field theory).  Because they
are protected by supersymmetry, we expect that the supersymmetric
nonabelian brane configurations we have considered in Sections 2 and 3
should still be valid in the full theory when $\alpha'$ corrections
are included.

Another interesting question is how the configurations we have
constructed here behave under S-duality.  In general, we only expect
S-duality to hold at the quantum level.  For example, the quantization
of fundamental strings in the string networks considered in Section 2
only arises in the quantum theory.  While we expect that there are
string networks like the box configuration shown in
Figure~\ref{boxfig} which are invariant under S-duality and an
exchange of the $x$ and $y$ coordinates, this is probably not possible
to see in the classical theory.  On the other hand, some aspects of
S-duality may be visible even classically.  This is well-understood in
the abelian theory, where S-duality is equivalent to Hodge duality on
the field strength, but is not well understood in the supersymmetric
theory.  If one performs an S-duality on the configuration of
Fig.~\ref{juncorigfig}, after performing T-duality along the
transverse directions $x_2$ and $x_3$, one finds a brane configuration
in which D-strings and F-strings are replaced by D3-branes and smeared
D-strings respectively.  A Yang-Mills representation of such a
configuration is known \cite{Lee}; in this construction an
off-diagonal zero mode is present, which is similar to our deformation
parameter $a$. It would be interesting to investigate further these
nontrivial classical configurations which seem to be S-dual to each
other.

There is a close connection between the nonabelian currents we have
used in this paper and the Seiberg-Witten map \cite{SW}, which relates
noncommutative Yang-Mills theory to Born-Infeld theory.  The D-brane
multipole currents used here were applied in \cite{SWe,OO,MS,LM} to
obtain an explicit expression for the Seiberg-Witten map.  Computing
the Seiberg-Witten map for given matrix ($\sim$ noncommutative) field
configurations is equivalent to finding the RR current densities
$j^{0i}$ for these configurations.  For example, in \cite{KO} the
Seiberg-Witten map of fluxons \cite{fluxon,fluxon2,fluxon3,fluxon4}
showed tilted D-string configurations \cite{AkiKoji, uni} specific to
noncommutative monopoles \cite{AkiKoji, ncmono,GN}, while a
perturbative version of this phenomena has been observed in
\cite{HH}. It would be interesting to compute other components such as
$j^{0y}$ in these configurations to confirm this brane interpretation.

In this paper we have restricted attention only to the leading
short-time behavior of the intersecting brane tachyon.  The full
time-dependent tachyon condensation process may have interesting
applications to brane cosmology \cite{intersection-cosmology}. 
It would be interesting to investigate possible applications of the
methods developed in this paper to intersecting brane cosmological models.

%%%%%%%%%%%%%%%%%%%%%%%%%%%%%%%%%%%%%%%%%%%%%%%%%%%%%%%%%%%%%%%%%%%%%
%%%%%%%%%%%%%%%%%%%% Acknowledgements %%%%%%%%%%%%%%%%%%%%%%%%%%%%%%%
%%%%%%%%%%%%%%%%%%%%%%%%%%%%%%%%%%%%%%%%%%%%%%%%%%%%%%%%%%%%%%%%%%%%%

\acknowledgments{

We would like to thank H.~Ooguri, A.\ Sen, D.~Tong, J.~Troost, and
M.\ Van Raamsdonk
for helpful discussions and comments. 
K.~H.~ is grateful to the CTP at MIT for kind hospitality,
where a part of this work was done.

This work was supported in part by the Grant-in-Aid for Scientific
Research (No.~12440060, 13135205 and 15740143) from the Japan Ministry
of Education, Science and Culture and in part by the DOE through
contract $\#$DE-FC02-94ER40818.
}

%%%%%%%%%%%%%%%%%%%%%%%%%%%%%%%%%%%%%%%%%%%%%%%%%%%%%%%%%%%%%%%
%%%%%%%%%%%%%%%%%%%%%%%%%%%%%%%%%%%%%%%%%%%%%%%%%%%%%%%%%%%%%%%
%%%%%%%%%%%%%%%%%%%%%%%%%%%%%%%%%%%%%%%%%%%%%%%%%%%%%%%%%%%%%%%

%%%%%%%%%% References %%%%%%%%%%%%%%%%%%%%%%%%%
\newcommand{\J}[4]{{\sl #1} {\bf #2} (#3) #4}
\newcommand{\andJ}[3]{{\bf #1} (#2) #3}
\newcommand{\AP}{Ann.\ Phys.\ (N.Y.)}
\newcommand{\MPL}{Mod.\ Phys.\ Lett.}
\newcommand{\NP}{Nucl.\ Phys.}
\newcommand{\PL}{Phys.\ Lett.}
\newcommand{\PR}{ Phys.\ Rev.}
\newcommand{\PRL}{Phys.\ Rev.\ Lett.}
\newcommand{\PTP}{Prog.\ Theor.\ Phys.}
\newcommand{\hep}[1]{{\tt hep-th/{#1}}}
%%%%%%%%%%%%%%%%%%%%%%%%%%%%%%%%%%%%%%%%%%%%%%%

\end{document}